\documentclass[reprint,amsmath,amssymb,amsart,aps,superscriptaddress,twocolumn,float]{revtex4-2}
\usepackage{graphicx,graphics}
\usepackage{dcolumn}
\usepackage{bm}
\usepackage{amssymb}
\usepackage{amsfonts}
\usepackage{float}
\usepackage{xcolor}
\usepackage[libertine]{newtxmath}

\begin{document}

\title{Magnetism-induced second-order nonlinear optical responses in multiferroic BiFeO$_3$} 	

\author{Babu Baijnath Prasad} 
\affiliation{Department of Physics and Center for Theoretical Physics, National Taiwan University, Taipei 10617, Taiwan, Republic of China\looseness=-1}
\affiliation{Nano Science and Technology Program, Taiwan International Graduate Program, Academia Sinica, Taipei 11529, Taiwan, Republic of China\looseness=-1}

\author{Guan-Fu Liu} 
\affiliation{Department of Physics and Center for Theoretical Physics, National Taiwan University, Taipei 10617, Taiwan, Republic of China\looseness=-1}

\author{Guang-Yu Guo} 
\email{gyguo@phys.ntu.edu.tw} 
\affiliation{Department of Physics and Center for Theoretical Physics, National Taiwan University, Taipei 10617, Taiwan, Republic of China\looseness=-1}
\affiliation{Physics Division, National Center for Theoretical Sciences, Taipei 10617, Taiwan, Republic of China\looseness=-1}

\date{\today}

\begin{abstract}
Nonlinear optical (NLO) responses of noncentrosymmetric nonmagnets have drawn a lot of attention in the past decades 
because of their significance in materials characterization, green energy and device applications.
On the other hand, magnetism not only can break the inversion symmetry in centrosymmetric crystals but also introduce novel NLO phenomena 
in noncentrosymmetric materials, thus enabling the magnetic field control of light-matter interactions.
However, the magnetism-induced NLO responses have rarely been studied so far.
In this paper, we first extend the numerical calculation friendly formula by Rashkeev {\it et al.} 
[Phys. Rev. B {\bf 57}, 3905 (1998)] for second harmonic generation (SHG) in nonmagnetic 
materials to include magnetic systems and then calculate the magnetism-induced NLO responses of bismuth ferrite (BiFeO$_3$), 
a multiferroic that exhibits both ferroelectricity and antiferromagnetic (AFM) ordering
at room temperature and has a band gap that falls in the visible frequency region.
First, we find that the calculated magnetism-induced SHG susceptibilities are large and the SHG intensity 
is tunable with the reversal of magnetization. 
In particular, the interference between crystallographic SHG and magnetically induced SHG components
results in a strong magnetic contrast of the SHG signal of $\sim$440\% at SHG photon energy of 4.82 eV,
thus enabling a magnetic control of the SHG in multiferroic BiFeO$_3$. 
Also, because of the sensitivity of the SHG signal to the direction of the N{\'e}el vector, the SHG can be utilized to detect 
the reversal of the N{\'e}el vector in the AFM materials, which is an important issue for AFM spintronics.
Second, the calculated bulk photovoltaic effects (BPVE) in BiFeO$_3$ are also strong, being larger than 
some well-known NLO compounds such as BaTiO$_3$, GaAs, CdS and CdSe.
Finally, we analyse the origins of the prominent features in the NLO response spectra 
in terms of the calculated quantum geometric quantities.
Our interesting findings suggest that the magnetism-driven NLO responses in BiFeO$_3$ are significant, 
anisotropic and tunable, and that understanding the magnetism-driven components of both SHG and BPVE 
is essential for their applications in, e.g., multiferroic-based photovoltaic devices and second-harmonic generation.
\end{abstract}

\maketitle

\section{INTRODUCTION}

Strong second-order nonlinear optical (NLO) responses could be produced by materials lacking inversion ($P$) symmetry under intense optical fields \cite{Boyd2003,Shen2003}. 
Second-harmonic generation (SHG) \cite{Franken1961,Kleinman1962} and bulk photovoltaic effect (BPVE) \cite{Glass1974,Koch1975,Kraut1979,Baltz1981,Fridkin2001} 
are two well-known examples of second-order NLO responses. 
SHG, a specific instance of sum-frequency generation in noncentrosymmetric crystals, is widely used as a tool for symmetry characterization and frequency doublers \cite{Shen2003}.
Since the 1960s, the SHG has been studied extensively in bulk semiconductors
\cite{Chang1965,Levine1991a,Levine1991b,Sipe1993,Hughes1996,Cai2009,Cheng2019,Song2020a}, and more recently also in one-dimensional \cite{Guo2004,Guo2005,Wu2008} 
and two-dimensional \cite{Wang2015,Hu2017,Wang2017,Attaccalite2019,Song2020b,Gudelli2020,Cheng2021,Gudelli2021} materials.
However, the study of magnetism-induced SHG elements has only received 
attention lately (see, e.g., Refs. \cite{Fiebig2005,Chen2022,Xiao2022} and references therein).
These magnetism-induced SHG can be used as a tool to probe surface and interface magnetization. 
This could also be used to manipulate the SHG by a magnetic field, as shown recently by Toyoda {\it et al.}~\cite{Toyoda2023}, where stronger SOC enhances the SHG contribution by the magnetic order. 
This results in comparable magnitudes for both crystalline and magnetic SHGs, leading to novel phenomena such as magnetic switching of SHG \cite{Toyoda2023}. 

Another intriguing second-order NLO response is the generation of dc photocurrents, commonly known as the bulk photovoltaic effect or photogalvanic effect.
It emerges from the inversion-asymmetric transition of electron position or velocity during the optical excitation, 
and the resulting dc photocurrents are, respectively, called the shift current and the injection current \cite{Sipe2000}.
In nonmagnetic systems, shift currents are produced by linearly polarized light, while injection currents are produced by circularly polarized light, 
and often called linear shift and circular injection currents. 
However, linearly (circularly) polarized light can produce injection (shift) currents as well as shift (injection) currents 
due to time-reversal ($T$) symmetry breaking in magnetic systems \cite{Ahn2020}.
For instance, in $PT$-symmetric systems, only the linear injection and circular shift current responses appear.
Nonetheless, when both $T$ and $PT$ symmetries are broken, all four responses, namely, linear (circular) shift and linear (circular) injection currents can appear.
LiNbO$_3$ \cite{Glass1974}, BaTiO$_3$ \cite{Koch1975,Young2012a}, and PbTiO$_3$ \cite{Young2012a} 
are the first few well-known examples of ferroelectric oxides in which BPVE has been extensively investigated.
Recently, there has been a resurgence of interest in studying the bulk photovoltaic response in magnetic systems 
(see, e.g., Refs. \cite{Gudelli2020,Ahn2020,Dai2022} and references therein). 
A few examples include the $PT$-symmetric antiferromagnetic (AFM) bilayer CrI$_3$ \cite{Zhang2019,Gudelli2020} 
and MnBi$_2$Te$_4$ \cite{Fei2020,Wang2020}, AFM Dirac semimetal MnGeO$_3$ \cite{Ahn2020}, as well as ferromagnetic Weyl semimetal PrGeAl 
(neither $T$ nor $PT$ symmetry) \cite{Ahn2020}.  
In these topological semimetals, the divergent behavior of shift and injection current conductivities 
at low frequencies has been attributed to the corresponding divergence in the quantum geometric quantities \cite{Ahn2020}. 
Furthermore, the circular photogalvanic current can be utilized to measure the topological charge (i.e. Chern number) 
of Weyl or higher-order nodes by optical means \cite{Ma2017,Juan2017,Hsu2023}.
Thus, second-order dc photocurrents can serve as a novel and effective tool for experimentally investigating 
the quantum geometry in materials \cite{Ahn2020,Ahn2022}. 
Additionally, magnetism-induced circular shift and linear injection current can be used to distinguish between 
different magnetic phases since they are directly connected with the magnetic point group of the crystal, 
carrying unique features for various magnetic structures (symmetries) \cite{Pi2023}. 
Also, by tuning the material close to the magnetic phase transition (e.g., by applying a large electric field), 
one can break certain symmetries present in the crystal, which can give rise to new BPVE tensor elements 
or change the magnitude (sign) of already existing responses. 
As a result, a critical enhancement of certain responses can be expected.
It is important to note that large bulk photovoltaic materials 
could also be used in cutting-edge solar-cell designs \cite{Butler2015,Tan2016,Cook2017}.
  
In recent years, ferroelectric materials have become a popular choice 
for investigating second-order NLO responses because of its noncentrosymmetric structure.
Bismuth ferrite (BiFeO$_3$) is one of the most extensively researched multiferroics 
because it is one of the few compounds that exhibit magnetic order and ferroelectricity in the same phase at ambient temperature.
It belongs to the class of Type I multiferroics where the coexisting orders come from independent mechanisms \cite{Sousa2016}.
For instance, at $T < 1093$ K, BiFeO$_3$ becomes ferroelectric, and only at far lower temperatures (below 643 K), 
it becomes antiferromagnetic ($G$-type AFM structure). 
This clearly suggests that the mechanisms governing magnetism and ferroelectricity are totally distinct from one another.  
Also, it possesses a large spontaneous ferroelectric polarization of 90-100 $\mu$C/cm$^2$ along the [111] direction \cite{Neaton2005}.
Compared with many typical ferroelectric oxides (band gap $\approx$ 3.5 eV), BiFeO$_3$ 
has a relatively narrow direct optical band gap of about 2.74 eV which lies in the visible spectrum \cite{Ihlefeld2008}.
As a result, it receives considerable interest as a promising material for ferroelectric-based photovoltaic devices \cite{Yang2010,Grinberg2013}.

In 2008, large optical SHG coefficients in thin films of BiFeO$_3$ were measured by Kumar {\it et al.} \cite{Kumar2008}.
Later on, Ju {\it et al.} performed the GGA+$U$ calculations of the linear dielectric function and second-harmonic generation in BiFeO$_3$ \cite{Ju2009}. 
However, Ju {\it et al.} used a formalism which is valid for nonmagnetic materials only and they did not consider the effect of relativistic spin-orbit coupling (SOC).
Consequently, magnetism-induced SHG components studied in this paper, would not appear in their GGA+$U$ calculations \cite{Ju2009}.
Recently, Xu {\it et al.} reported a magnetoelectric coupling in BiFeO$_3$ films probed by external magnetic fields and wide temperature-range SHG, 
showing the ability of the magnetic field to control the nonlinear polarization caused by light \cite{Xu2023}. 
It would be important to investigate the magnetism-induced SHG components, which have been overlooked so far,  because they could give rise to a number
of novel phenomena such as switching of the SHG by a magnetic field (see, e.g., Ref. \cite{Toyoda2023} and references therein).

Bulk crystals of BiFeO$_3$ have also been found to have a switchable-diode effect and a visible-light photovoltaic effect \cite{Choi2009}. 
Yang {\it et al.} \cite{Yang2009} reported the photovoltaic effect in BiFeO$_3$ thin films with external quantum efficiencies up to $\sim$10\%.
Subsequently, above-bandgap voltages at ferroelectric domain walls in BiFeO$_3$ thin films were observed \cite{Yang2010,Ji2010}.
Seidel {\it et al.} studied the BPVE in ferroelectric BiFeO$_3$ thin films with periodic domain structures and found out that ferroelectric domain walls act as current sources \cite{Seidel2011}.
Moreover, in BiFeO$_3$ thin films, it is found that the bulk photovoltaic tensor coefficient $\beta_{22}$ 
is approximately five orders of magnitude larger than that of other conventional ferroelectric materials in the visible range of the solar spectrum \cite{Ji2011}.
Alexe {\it et al.} studied the anomalous photovoltaic effect in BiFeO$_3$ single crystals 
and their results have shown that the external quantum efficiency can be further enhanced by up to seven orders of magnitude using a nanoscale top electrode \cite{Alexe2011}. 
Using first-principles calculation, Young {\it et al.} investigated the nonmagnetic linear shift current tensor elements of BiFeO$_3$ \cite{Young2012b}. 
Very recently, Knoche {\it et al.} \cite{Knoche2021} experimentally studied the circular BPVE in epitaxially grown BiFeO$_3$ thin films with stripe-domain pattern.  
Nevertheless, the magnetism-induced bulk photovoltaic effect, namely, circular shift current and linear injection current, have not been investigated so far in BiFeO$_3$. 
Additionally, there is currently a lack of theoretical study on circular injection current in BiFeO$_3$.

In this paper, therefore, we present a systematic study of both 
structural and magnetic SHG and BPVE tensor elements of multiferroic BiFeO$_3$ by performing {\it ab initio} density functional theory calculations.
The rest of this paper is organized as follows.
In Sec. II, we present the crystal structure of BiFeO$_3$ along with the computational details 
for calculating the magnetism-induced SHG susceptibility and bulk photovoltaic effect. 
The main results are presented in Secs. III and IV.
In Sec. III, we first present magnetism-induced SHG, and then  
show the tunability in the SHG intensity with the reversal of magnetization direction.
Calculated magnetism-induced bulk photovoltaic responses are presented in Sec. IV.
In Sec. IV, we also present the calculated quantum geometric quantities to understand the features in these bulk photovoltaic spectra.
Furthermore, we present the calculated crystallographic NLO properties (i.e., the $i$-type SHG susceptibilities, linear shift current 
and circular injection current conductivities) in the Supplementary Material.
Finally, the conclusions drawn from this work are summarized in Sec. V.

\section{THEORY AND COMPUTATIONAL DETAILS}

\begin{figure}[htbp] \centering
\includegraphics[width=7.5cm]{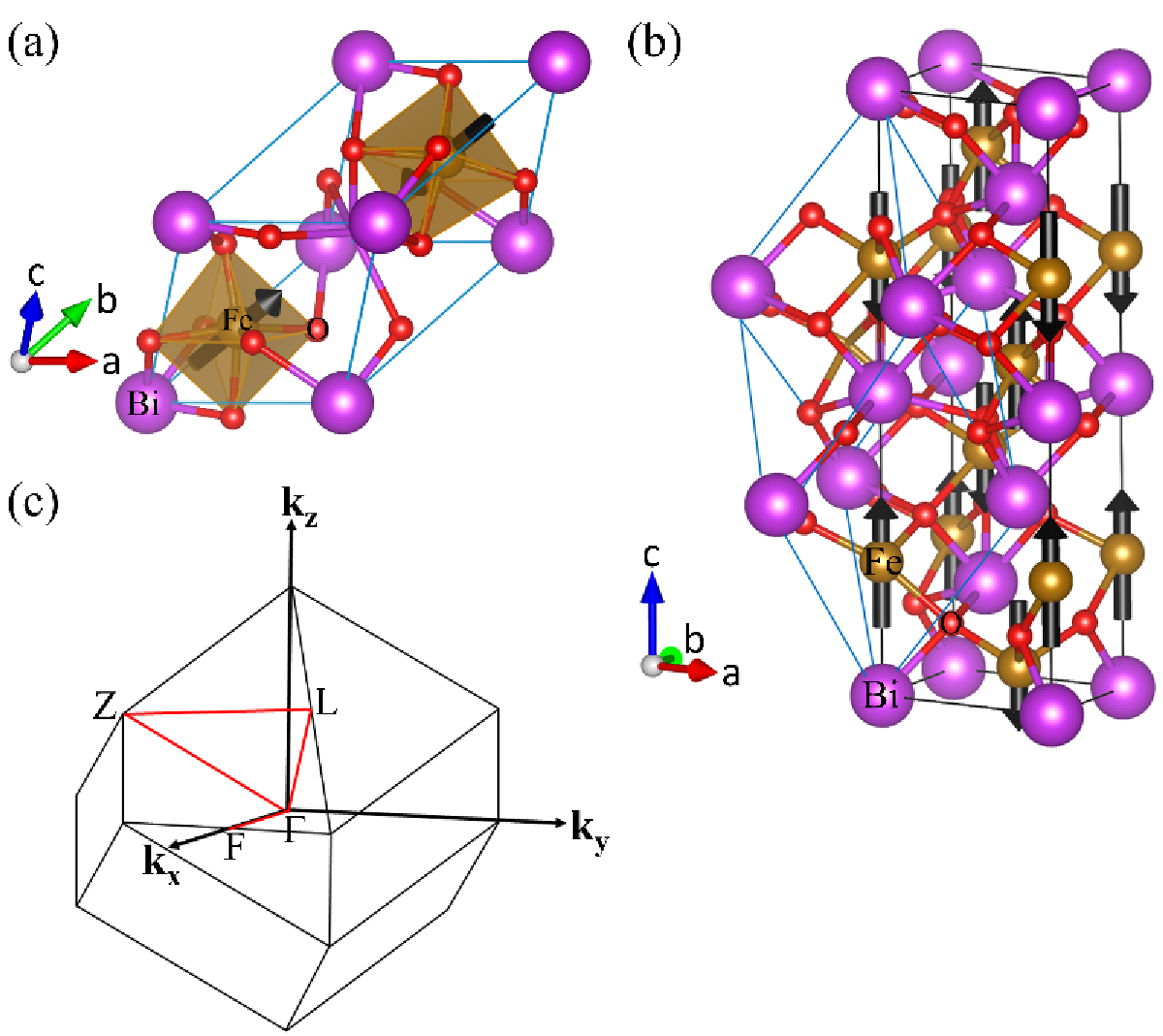}
\caption{Crystal and magnetic structure of BiFeO$_3$. (a) Rhombohedral primitive unit cell. (b) Hexagonal conventional unit cell showing the G-type AFM configuration, 
with magnetic moments on Fe atoms denoted by black arrows. The solid blue lines denote the rhombohedral primitive unit cell. (c) The corresponding Brillouin zone.}
\label{fig:crystal}
\end{figure}

BiFeO$_3$ has a polar structure with trigonal space group {\it R}$3${\it c} at room temperature (see Fig. \ref{fig:crystal}).
Figure \ref{fig:crystal}(a) and \ref{fig:crystal}(b) show the rhombohedral primitive unit cell 
and the hexagonal conventional unit cell of BiFeO$_3$, respectively.
Such a polar structure results from the counter-rotations of nearby oxygen octahedra about the threefold [111] axis, 
making the spontaneous polarization along the [111] direction feasible \cite{Neaton2005}.
Also, along this threefold axis, Bi, Fe and O atoms are displaced from each other.
It is well-known that Fe atoms in BiFeO$_3$ form a $G$-type antiferromagnetic ordering as shown in Figs. \ref{fig:crystal}(a)-\ref{fig:crystal}(b). 
Note that [111] direction in the primitive unit cell corresponds to the [001] direction ($z$ direction) in the hexagonal conventional unit cell. 
The experimentally determined atomic positions and lattice constants~\cite{Kubel1990} are used in the present study. 

Our self-consistent electronic structure calculations are based on the density functional theory 
with the generalized gradient approximation (GGA) \cite{Perdew1996}.
The accurate projector augmented-wave method \cite{Blochl1994}, 
as implemented in the Vienna Ab Initio Simulation Package (VASP) \cite{Kresse1993,Kresse1996}, is used. 
To better describe the on-site Coulomb interaction among the Fe 3$d$ electrons, we adopt the Dudarev's GGA+$U$ scheme \cite{Dudarev1998}.
Previous studies suggest that effective $U$ should be between 4-6 eV \cite{Ju2009,Young2012b}.
Thus we use $U_{eff} = 5$ eV here.
The valence electronic configurations of Bi, Fe, and O adopted here 
are $5d^{10}6s^26p^3$,  $3p^63d^74s^1$, and $2s^22p^4$, respectively.
Unlike the previous study~\cite{Ju2009}, the SOC is included in the present work because it causes the magnetism-induced NLO responses.
A large plane wave energy cutoff of 500 eV is used throughout.
In the self-consistent electronic structure calculations, a $\Gamma$-centered $k$-point 
mesh of 10 $\times$ 10 $\times$ 10 is used in the Brillouin zone (BZ) integration 
by the tetrahedron method \cite{Jepson1971,Temmerman1989}.
The calculated band gap of 2.40 eV is indirect. Nevertheless, the direct band gap of 2.43 eV is only
slightly larger than the indirect one and agrees quite well with the experimental direct optical band gap of 2.74 eV \cite{Ihlefeld2008}.

We then calculate the nonlinear optical responses from the calculated relativistic band structures 
within the linear response formalism with the independent-particle approximation.
Using the length gauge formalism, Aversa and Sipe presented a general formalism for nonlinear optical response calculations \cite{Aversa1995}.
Subsequently, by taking the time-reversal symmetry and also the symmetry under permutation of the indices into account, 
Rashkeev \textit{et al.} \cite{Rashkeev1998} derived a numerical calculation friendly formula for the SHG susceptibility $\chi^{(2)}$
of nonmagnetic materials. Following Rashkeev \textit{et al.} \cite{Rashkeev1998}, here we extend this formalism to include the magnetic systems 
(see Note 3 of the Supplemental Material (SM) \cite{Prasad-SM} for derivation).
The SHG susceptibility $\chi_{abc}^{(2)}(-2 \omega; \omega, \omega)$ for a magnetic material can be written as, 
\begin{equation}
\label{eq:3}
\begin{aligned}
\chi_{abc}^{(2)}(-2 \omega; \omega, \omega)= \chi_{abc,e}^{(2)}(-2 \omega; \omega, \omega)+\chi_{abc,i}^{(2)}(-2 \omega; \omega, \omega),  
\end{aligned}
\end{equation}
where
\begin{equation}
\label{eq:4}
\begin{aligned}
\chi_{abc,e}^{(2)}=& \frac{-e^3}{2 \varepsilon_0 \hbar^2} \int \frac{d^3 k}{(2 \pi)^3} \sum_{n m l} \frac{r_{n m}^a (r_{m l}^b r_{l n}^c+r_{m l}^c r_{l n}^b)}{\omega_{l n}-\omega_{m l}} \\
&\times [\frac{2 f_{n m}}{\omega_{m n}-2 \omega}+\frac{f_{l n}}{\omega_{l n}-\omega}+\frac{f_{m l}}{\omega_{m l}-\omega}],
\end{aligned}
\end{equation}
is the contribution of the purely interband processes, and
\begin{equation}
\label{eq:5}
\begin{aligned}
\chi_{abc,i}^{(2)}=& \frac{-i e^3}{2 \varepsilon_0 \hbar^2} \int \frac{d^3 k}{(2 \pi)^3} \sum_{n m} f_{n m} [\frac{2 r_{n m}^a (r_{m n ; c}^b+r_{m n ; b}^c)}{\omega_{m n}(\omega_{m n}-2 \omega)} \\ &+ \frac{r_{n m ; c}^a r_{m n}^b+r_{n m ; b}^a r_{m n}^c}{\omega_{m n}(\omega_{m n}-\omega)} \\
&+ \frac{r_{n m}^a(r_{m n}^b \Delta_{m n}^c+r_{m n}^c \Delta_{m n}^b)}{\omega_{m n}^2} (\frac{1}{\omega_{m n}-\omega}-\frac{4}{\omega_{m n}-2 \omega}) \\
&- \frac{r_{n m ; a}^b r_{m n}^c+r_{n m ; a}^c r_{m n}^b}{2 \omega_{m n}(\omega_{m n}-\omega)} \\
&+ \frac{\Delta_{m n}^a(r_{n m}^b r_{m n}^c+r_{n m}^c r_{m n}^b)}{4 \omega_{m n}^{2}(\omega_{m n}-\omega)}],
\end{aligned}
\end{equation}
is the contribution of the mixed interband and intraband processes.
Here $a$, $b$ and $c$ denote Cartesian directions.
Also, we assume here that $e > 0$ and the electron charge is $-e$. 
The only difference between the magnetic and nonmagnetic systems is the last term in Eq. \ref{eq:5} 
which vanishes in the presence of time-reversal symmetry. 
Here $r_{n m}^a$ and $r_{n m ; b}^a$ are the $a$-component of the interband position matrix element 
and its generalized momentum derivative, respectively.
$\Delta_{n m}^a$ is the difference between the electronic velocities at the bands $n$ and $m$.
$f_{n m} = f(\varepsilon_{n \mathbf{k}})-f(\varepsilon_{m \mathbf{k}})$ is the difference of the Fermi distribution functions. 
$\omega_{n m} = (\varepsilon_{n \mathbf{k}}-\varepsilon_{m \mathbf{k}})/\hbar$ where $\varepsilon_{n \mathbf{k}}$ is the $n$th band energy at the $\textbf{k}$ point, and $\varepsilon_0$ is the vacuum permittivity. 
We notice that this formula has also been derived independently earlier by Chen {\it et al.} \cite{Chen2022}.
In the present calculations, we replace $\omega$ by ($\omega+ i\frac{\eta}{\hbar}$) 
where $\eta$ is a fixed smearing parameter and we use $\eta$ = 0.04 eV.

The nonzero elements of the SHG susceptibility tensor of a magnetic material are usually divided into two types, namely, 
nonmagnetic $i$-type ($\chi^{(i)}$) due to the structural asymmetry and magnetic $c$-type ($\chi^{(c)}$) 
due to the broken time-reversal symmetry (i.e., magnetism) \cite{Fiebig2005},
\begin{equation}
\label{eq:9}
\begin{aligned}
\chi_{abc}^{(2)}= \chi_{abc}^{(i)}+\chi_{abc}^{(c)}.  
\end{aligned}
\end{equation}
Since the time-reversal operation on $\chi_{abc}^{(2)}$ is equivalent to taking the complex conjugate of it, 
here we simply obtain $\chi_{abc}^{(i)}$ and $\chi_{abc}^{(c)}$ as the symmetrized and anti-symmetrized parts of $\chi_{abc}^{(2)}$, respectively.
We note that below the mid-band gap, the SHG susceptibility is purely real. 
Thus, the imaginary part of both $i$-type and $c$-type SHG is zero below half of the band gap. 
Furthermore, due to their anti-symmetric nature, the real part of $c$-type SHG elements should also be zero. 
As a result, only the real part of $i$-type SHG persists below the mid-band gap.
Interestingly, we note that $i$-type SHG is independent of magnetization direction whereas $c$-type SHG changes sign 
when the magnetization direction is reversed. 
This formalism was recently applied to calculate the SHG spectra of ferrimagnetic Eu$_2$MnSi$_2$O$_7$ 
and the results nicely explained the observed magnetic-field switching of SHG~\cite{Toyoda2023}. 

Another interesting second-order nonlinear optical response in a noncentrosymmetric material is
the generation of dc photocurrents~\cite{Sipe2000,Nastos2006,Nastos2010,Azpiroz2018,Ahn2020}.
The dc photocurrent density along the $a$-axis is given by~\cite{Sipe2000,Nastos2006,Nastos2010,Azpiroz2018,Ahn2020}
\begin{equation}
\label{eq:6}
\begin{aligned}
J_a(0)=\sum_{b c} \sigma_{a b c}(0 ; \omega,-\omega) E_b(\omega) E_c(-\omega),
\end{aligned}
\end{equation}  
where $E_b$ and $E_c$ are the applied optical electric fields.
The photocurrent conductivity $\sigma_{a b c}(0 ; \omega,-\omega)$ is a third-rank tensor which contains two main contributions, namely, shift current and injection current~\cite{Sipe2000,Nastos2010}, i.e., $\sigma_{a b c}=\sigma_{a b c}^{sh}+\sigma_{a b c}^{inj}$.
Here $\sigma_{a b c}^{sh}$ and $\sigma_{a b c}^{inj}$ are the shift and injection current conductivities, respectively. 
Also, $\sigma_{a b c}^{inj}=\uptau \eta_{a b c}$ where $\uptau$ and $\eta_{a b c}$ are the relaxation time of photoexcited carriers and injection current susceptibility, respectively.
Note that $\sigma_{a b c}^{sh}$ does not depend on $\uptau$. 
Within the length gauge formalism, the shift current conductivity ($\sigma_{a b c}$) and injection current susceptibility ($\eta_{a b c}$) for a magnetic material can be written as~\cite{Ahn2020},
\begin{equation}
\label{eq:7}
\begin{aligned}
\sigma_{a b c}= \frac{-i \pi e^3}{\hbar^2} \int \frac{d^3 k}{(2 \pi)^3} \sum_{n m} f_{n m} (r_{n m}^c r_{m n ; a}^b-r_{n m ; a}^c r_{m n}^b) \delta(\omega_{m n}-\omega),
\end{aligned}
\end{equation}
and 
\begin{equation}
\label{eq:8}
\begin{aligned}
\eta_{a b c}= \frac{-2 \pi e^3}{\hbar^2} \int \frac{d^3 k}{(2 \pi)^3} \sum_{n m} f_{n m} \Delta_{m n}^a r_{n m}^c r_{m n}^b \delta(\omega_{m n}-\omega).
\end{aligned}
\end{equation}

Since a large number of $k$ points are required to get accurate 
nonlinear optical responses (SHG and BPVE), we use the efficient Wannier function interpolation scheme \cite{Yates2007,Azpiroz2018} based on 
the maximally localized Wannier functions (MLWFs) \cite{Marzari2012} as implemented 
in the WANNIER90 package \cite{Pizzi2020}.
Total 68 MLWFs per unit cell of Bi $p$, Fe $d$ and O $p$ orbitals are constructed by fitting to 
the GGA+$U$+SOC band structure.
The calculated Wannier interpolated band structure is identical to that from the {\it ab initio} calculation, 
as can be seen in Fig. S1 in the SM \cite{Prasad-SM}.
The SHG susceptibility, shift current conductivity, injection current susceptibility, and quantum geometric quantities are then evaluated by taking 
a dense $k$ mesh of 100 $\times$ 100 $\times$ 100.
However, for the group velocity difference, a denser $k$ mesh of 300 $\times$ 300 $\times$ 300 is used.
Test calculations using several different sets of $k$ meshes show that these calculated spectra converge within a few percent.
 
\begin{table*}
\caption{Nonzero symmetry elements of the SHG susceptibility and BPVE tensors for BiFeO$_3$. 
$\chi_{abc}^{(i)}$ and $\chi_{abc}^{(c)}$ are $i$-type and $c$-type SHG susceptibilities, respectively. $\sigma_{abc}^{sh, L}$ ($\sigma_{abc}^{sh, C}$) is linear (circular) shift current conductivity whereas $\eta_{abc}^{inj, L}$ ($\eta_{abc}^{inj, C}$) is linear (circular) injection current susceptibility, respectively. 
Here $\chi_{abc}^{(i)}$, $\sigma_{abc}^{sh, L}$ and $\eta_{abc}^{inj, C}$ are due to structural asymmetry, whereas $\chi_{abc}^{(c)}$, $\sigma_{abc}^{sh, C}$ and $\eta_{abc}^{inj, L}$ are due to magnetism.  
Also, note that both the crystallographic and magnetic point group of BiFeO$_3$ is $3m$.} 
\begin{ruledtabular}
\begin{tabular}{c c c c c c}
$\chi_{abc}^{(i)}$ & $\chi_{abc}^{(c)}$ & $\sigma_{abc}^{sh, L}$ & $\sigma_{abc}^{sh, C}$ & $\eta_{abc}^{inj, L}$ & $\eta_{abc}^{inj, C}$ \\ \\ \hline
$ xxz = yyz $ & $ xxz = yyz $ & $ xxz = yyz $ & $ xxz = yyz = -xzx = -yzy $ & $ xxz = yyz $ & $ xxz = yyz = -xzx = -yzy $ \\ 
$ xxy = yxx = -yyy $ & $ xxy = yxx = -yyy $ & $ xxy = yxx = -yyy $ & & $ xxy = yxx = -yyy $ & \\
$ zxx = zyy $ & $ zxx = zyy $ & $ zxx = zyy $ & & $ zxx = zyy $ & \\
$ zzz $ & $ zzz $ & $ zzz $ & & $ zzz $ & \\ \\
\end{tabular}
\end{ruledtabular}
\label{table:1}
\end{table*}

\section{Second-harmonic generation}

The SHG susceptibility of a material is a third-rank tensor ($\chi_{abc}^{(2)}$; $a,b,c=x,y,z$), 
and thus has 27 tensor elements. 
However, since BiFeO$_3$ has a trigonal structure with crystallographic point group $3m$, 
it has only four independent nonzero elements, namely, $\chi_{xxy}^{(2)}$, $\chi_{xxz}^{(2)}$, $\chi_{zxx}^{(2)}$, 
and $\chi_{zzz}^{(2)}$ \cite{Gallego2019}.
Other nonzero elements are related to these four elements by $ \chi_{xxy}^{(2)} = \chi_{yxx}^{(2)} = -\chi_{yyy}^{(2)}$, $ \chi_{xxz}^{(2)} = \chi_{yyz}^{(2)}$, and $ \chi_{zxx}^{(2)} = \chi_{zyy}^{(2)} $ (see Table I).
These are called $i$-type SHG susceptibility tensor and are denoted as $\chi_{xxy}^{(i)}$, $\chi_{xxz}^{(i)}$, $\chi_{zxx}^{(i)}$, and $\chi_{zzz}^{(i)}$.
Interestingly, the magnetic point group of $G$-type AFM BiFeO$_3$ is also $3m$.
Thus, as for the $i$-type SHG tensor, there are four independent nonzero elements for the $c$-type SHG tensor, 
namely, $\chi_{xxy}^{(c)}$, $\chi_{xxz}^{(c)}$, $\chi_{zxx}^{(c)}$, and $\chi_{zzz}^{(c)}$ (Table I) \cite{Gallego2019}. 
Here we focus on the magnetism-induced $c$-type SHG susceptibilities and 
we present the $i$-type SHG susceptibilities in the Supplementary Note 1 of the SM \cite{Prasad-SM}.

\begin{figure}[htbp] \centering
\includegraphics[width=8.0cm]{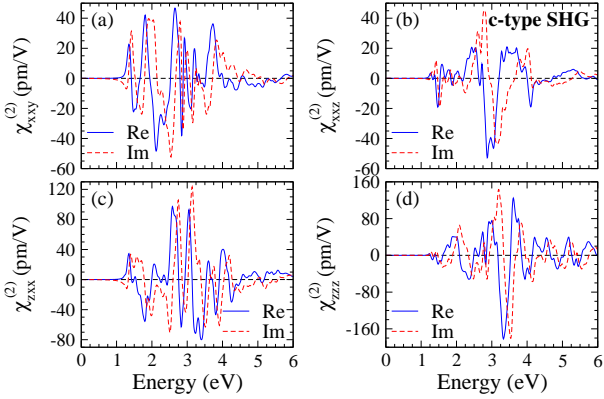}
\caption{Real (Re) and imaginary (Im) parts of the four independent nonzero $c$-type SHG susceptibility elements 
(a) $\chi_{xxy}^{(2)}$, (b) $\chi_{xxz}^{(2)}$, (c) $\chi_{zxx}^{(2)}$, and (d) $\chi_{zzz}^{(2)}$.}
\label{fig:shg-c-type}
\end{figure}

\subsection{Magnetism-induced second-harmonic generation}

\begin{figure}[htbp] \centering
\includegraphics[width=8.0cm]{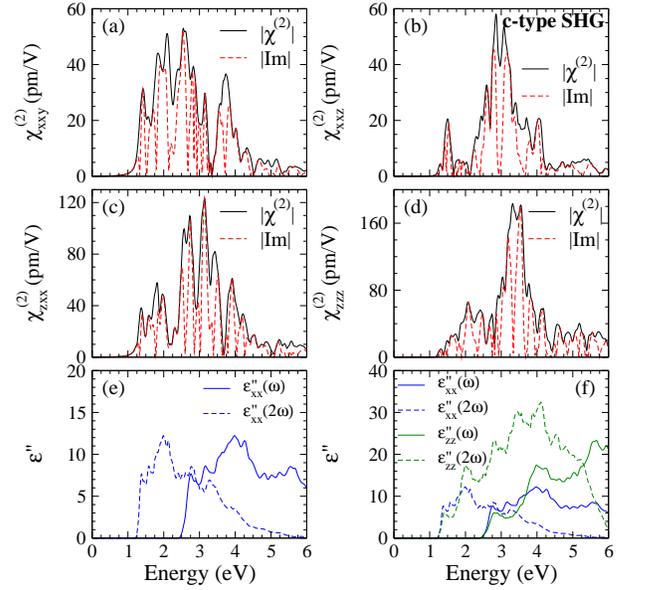}
\caption{Absolute values of nonzero $c$-type SHG susceptibility elements (a) $\chi_{xxy}^{(2)}$, (b) $\chi_{xxz}^{(2)}$, (c) $\chi_{zxx}^{(2)}$, and (d) $\chi_{zzz}^{(2)}$; as well as (e), (f) imaginary part of the dielectric function ($\varepsilon^{\prime\prime}$) of BiFeO$_3$.} 
\label{fig:shg-c-type-epsilon}
\end{figure}

Figure \ref{fig:shg-c-type} shows the real and imaginary parts of the $c$-type SHG susceptibilities. 
As discussed before, the real and imaginary parts of $c$-type SHG elements 
are zero below the half of the band gap ($\sim$1.2 eV). 
As the photon energy increases above the mid-band gap, both the real and imaginary parts start to oscillate 
between positive and negative values as compared with the corresponding $i$-type SHG elements 
(see Fig. \ref{fig:shg-c-type} and Fig. S3 in the SM \cite{Prasad-SM}). 
The absolute values of $c$-type SHG susceptibilities are shown in Fig. \ref{fig:shg-c-type-epsilon}.
Clearly, two prominent peaks appear in the $c$-type SHG susceptibility spectra that are close in magnitude.
The calculated values of these SHG susceptibilities are 53 (51), 58 (54), 123 (110) and 183 (182) pm/V 
for I (II) prominent peaks corresponding to $\chi_{xxy}^{(2)}$, $\chi_{xxz}^{(2)}$, $\chi_{zxx}^{(2)}$ 
and $\chi_{zzz}^{(2)}$, respectively, at 2.55 (2.10), 2.86 (3.08), 3.14 (2.73) and 3.32 (3.54) eV.
Similar to $i$-type SHG susceptibility, the largest magnitude is for $\chi_{zzz}^{(2)}$ [183 (pm/V)] 
among all $c$-type SHG susceptibilities. 
We note that the $c$-type SHG susceptibility in BiFeO$_3$ is about one order of magnitude smaller 
than that of the corresponding $i$-type SHG elements.
This is because in BiFeO$_3$, the heavy Bi atom predominantly contributes to bands located 8 eV 
below or 5 eV above the top of the valence band. 
Therefore, the optical transitions are dominated by the Fe $d$ and O $p$ orbitals, 
both of which exhibit a much smaller SOC strength compared to that of the Bi atom 
(see Fig. S2 in the SM \cite{Prasad-SM}).

In order to understand the origin of the prominent features in the calculated $c$-type $\chi^{(2)}$ spectra, 
we plot the modulus of the imaginary part as well as the absolute values of $c$-type SHG susceptibilities 
and compare them with the absorptive (imaginary) part of the dielectric function $\varepsilon(\omega)$ 
in Fig. \ref{fig:shg-c-type-epsilon}.
First, we can divide the whole SHG spectra into two regions: the first region is in between mid-band gap 
and the absorption edge, corresponding to double-photon ($2\omega$) resonance whereas the second region 
(above absorption edge) is a mix of both single-photon ($\omega$) and double-photon resonances.
Nonetheless, because of the magnetic origin of the $c$-type SHG susceptibility, there is no direct correlation 
between the $c$-type SHG elements and the imaginary part of the optical dielectric function 
($\varepsilon^{\prime\prime}$) [see Fig. \ref{fig:shg-c-type-epsilon}].
Figure \ref{fig:shg-c-type-epsilon} shows that only a few comparable peaks of the $c$-type SHG susceptibilities are visible.
Thus, we can conclude that, in comparison to $i$-type SHG, the link between $c$-type SHG susceptibility 
and the optical dielectric function is weaker.

\begin{figure}[htbp] \centering
\includegraphics[width=8.0cm]{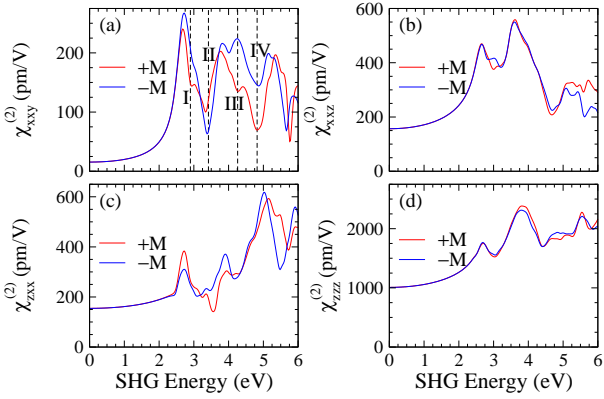}
\caption{Absolute values of nonzero SHG susceptibility elements (a) $\chi_{xxy}^{(2)}$, (b) $\chi_{xxz}^{(2)}$, (c) $\chi_{zxx}^{(2)}$, and (d) $\chi_{zzz}^{(2)}$ as a function of SHG photon energy for magnetization directions parallel (red spectra) and antiparallel (blue spectra) to the $z$-axis, respectively.
In (a), the black dashed lines marked as I, II, III and IV correspond to 2.9, 3.42, 4.26 and 4.82 eV, respectively.	
} 
\label{fig:shg-M}
\end{figure}

\begin{figure}[htbp] \centering
\includegraphics[width=8.0cm]{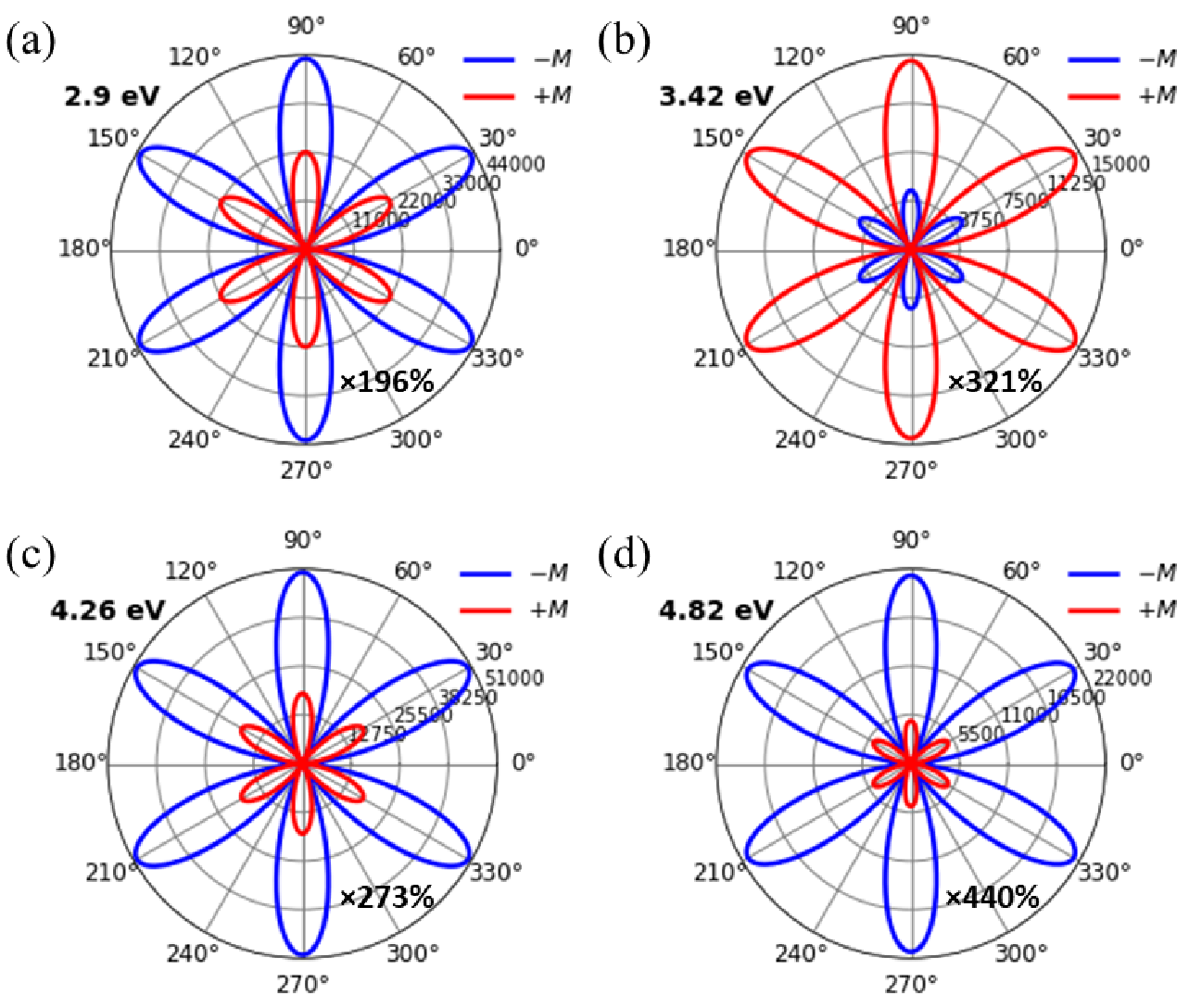}
\caption{Calculated in-plane polarization-resolved SHG pattern $|\chi_{\parallel}(\theta)|^2$ of BiFeO$_3$ as a function of azimuthal polarization for normal incidence at SHG photon energy of (a) 2.9 eV, (b) 3.42 eV, (c) 4.26 eV, and (d) 4.82 eV.
Note that the red and blue spectra are for magnetization direction parallel and antiparallel to the $z$-axis, respectively.} 
\label{fig:shg-z-in-pa}
\end{figure}

\begin{figure}[htbp] \centering
\includegraphics[width=8.0cm]{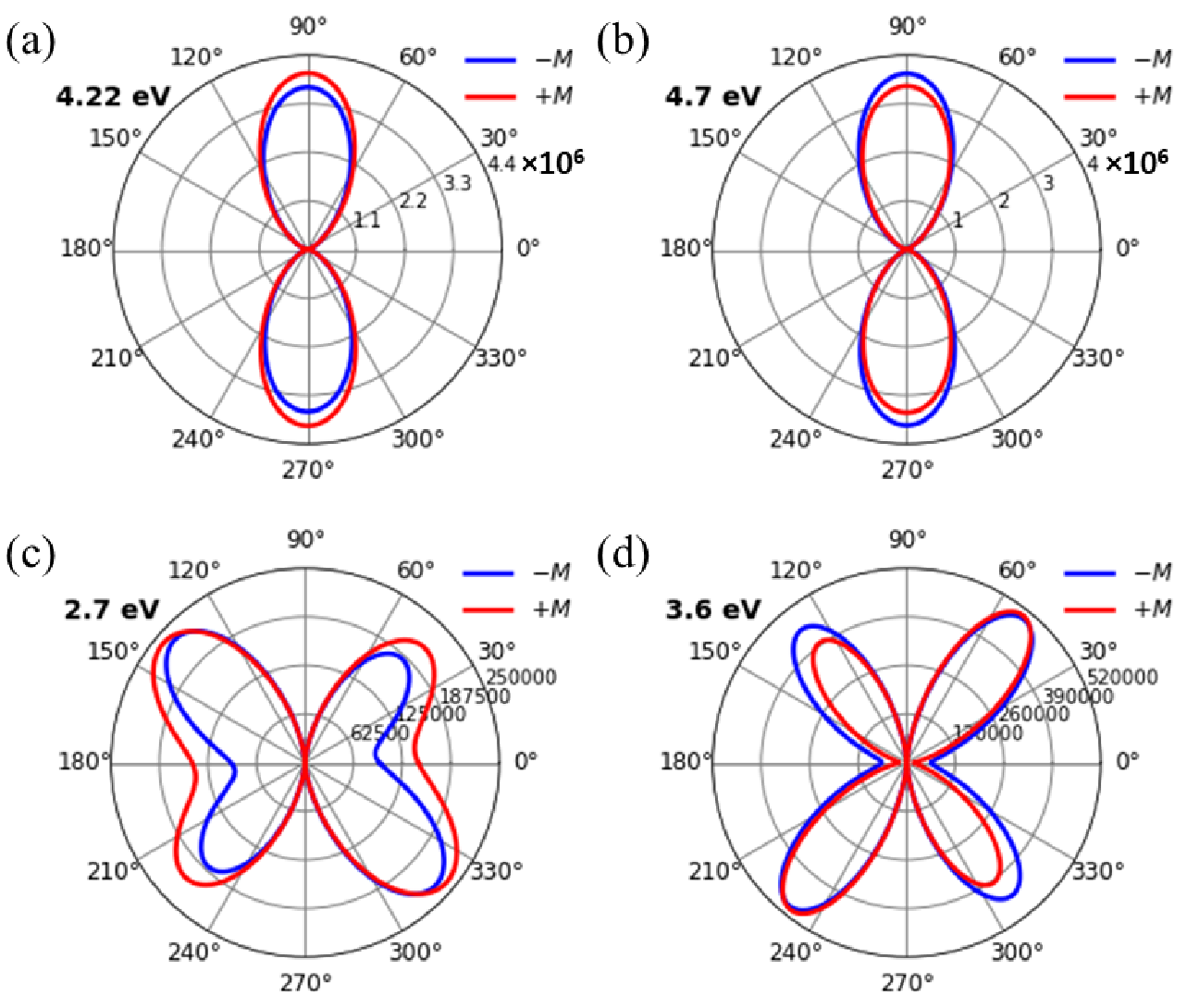}
\caption{Calculated in-plane polarization-resolved SHG pattern (a), (b) $|\chi_{\parallel}(\theta)|^2$, and (c), (d) $|\chi_{\perp}(\theta)|^2$ of BiFeO$_3$ as a function of azimuthal polarization for incident light along $x$-direction at SHG photon energy of (a) 4.22 eV, (b) 4.7 eV, (c) 2.7 eV, and (d) 3.6 eV.
Note that the red and blue spectra are for magnetization direction parallel and antiparallel to the $z$-axis, respectively.} 
\label{fig:shg-x-in-pa-pd}
\end{figure}

\subsection{Magnetization-direction tunable SHG pattern}

\begin{figure}[thbp] \centering
\includegraphics[width=8.0cm]{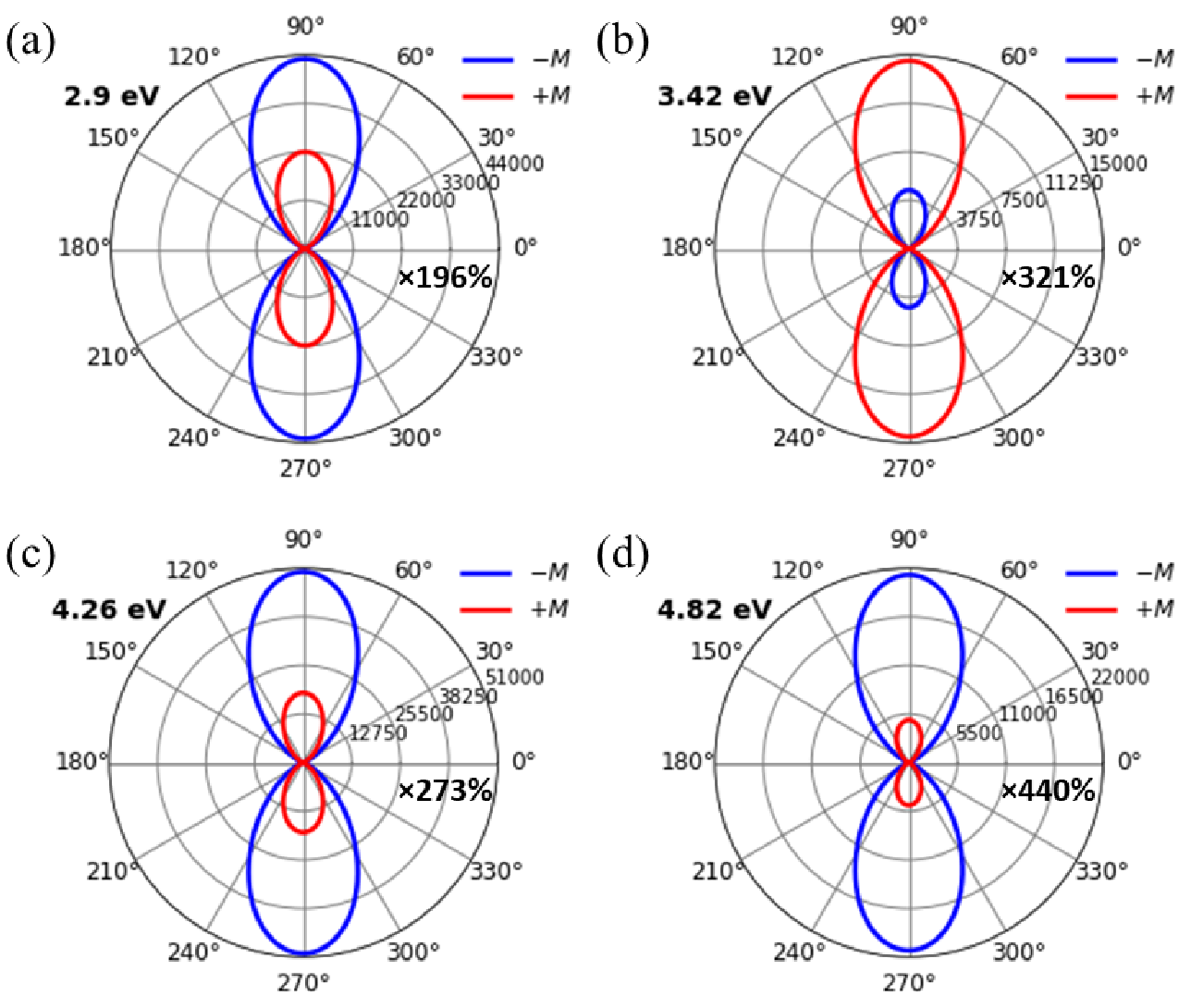}
\caption{Calculated out-of-plane polarization-resolved SHG pattern $|\chi_{out}(\theta)|^2$ of BiFeO$_3$ as a function of azimuthal polarization for incident light along $y$-direction at SHG photon energy of (a) 2.9 eV, (b) 3.42 eV, (c) 4.26 eV, and (d) 4.82 eV.
Note that the red and blue spectra are for magnetization direction parallel and antiparallel to the $z$-axis, respectively.} 
\label{fig:shg-y-out}
\end{figure}

Because BiFeO$_3$ is an antiferromagnet, it has no net magnetization. So, here the term magnetization means staggered magnetization (i.e., the N{\'e}el vector), which is the normalized difference of the unit vectors of the sublattice magnetizations. 
Recent studies~\cite{Wadley2016,Godinho2018,Grigorev2021,Zhang2022} have shown the electrical detection and reversal of the orientation of the N{\'e}el vector in antiferromagnetic materials, which makes them promising candidates for more reliable, quick and dense spintronic devices.
Figure \ref{fig:shg-M} depicts the absolute values of nonzero SHG susceptibilities as a function of SHG photon energy for magnetization directions parallel (red spectra) and antiparallel (blue spectra) to the $z$-axis, respectively.
It is evident that when magnetization direction is reversed, the spectra significantly differ. 
For instance, the absolute value of $\chi_{xxy}^{(2)}$ changes from 135 to 224 (pm/V), and 69 to 146 (pm/V) 
at SHG photon energy of 4.26 and 4.82 eV, respectively [label III and IV; Fig. \ref{fig:shg-M}(a)]. 
However, for $\chi_{xxz}^{(2)}$ and $\chi_{zzz}^{(2)}$, it is less pronounced for most of the considered SHG photon 
energy range [see Figs. \ref{fig:shg-M}(b) and \ref{fig:shg-M}(d)]. 
Since the SHG intensity is proportional to the square of the absolute value of SHG susceptibility, 
i.e., $I(2\omega)\propto|\chi_{a b c}^{(2)}(2\omega)|^{2}$, changes in SHG intensity are 
more prominent upon magnetization reversal, as discussed in detail below. 
In Fig. \ref{fig:shg-M}(a), the black dashed lines labeled I, II, III and IV correspond to 
Figs. \ref{fig:shg-z-in-pa}(a)-\ref{fig:shg-z-in-pa}(d) as well as Figs. \ref{fig:shg-y-out}(a)-\ref{fig:shg-y-out}(d), 
respectively, since these polarization-resolved SHG patterns depend solely on $\chi_{xxy}^{(2)}$ 
[see Eqs. (\ref{eq:14}) and (\ref{eq:16})]. 

Here we present the polarization-resolved SHG response of BiFeO$_3$.
Polarization-resolved SHG is a well-known characterization technique for probing the different magnetic symmetries in solids.
The photoinduced nonlinear polarization can be expressed as,
\begin{equation}
\label{eq:10}
\begin{aligned}
P_a(2\omega)=\varepsilon_{0} [\chi_{a b c}^{(i)}+\chi_{a b c}^{(c)}(\hat{M})] E_b(\omega) E_c(\omega),
\end{aligned}
\end{equation}  
where $E(\omega)$ is the electric field of the incident light and $\hat{M}$ is the magnetization unit vector. 
As mentioned earlier, $\chi_{a b c}^{(2)}$ consists of $i$-type ($\chi_{a b c}^{(i)}$) and $c$-type ($\chi_{a b c}^{(c)}$) SHG. 
Furthermore, unlike $\chi_{a b c}^{(i)}$, which is unaffected by the magnetization direction, 
$\chi_{a b c}^{(c)}$ depends on the magnetization direction ($\hat{M}$) and changes sign when 
$\hat{M}$ is reversed. Thus, the interference between these two contributions upon the reversal 
of magnetization direction results in a large change in the SHG intensity which is defined as,
\begin{equation}
\label{eq:11}
\begin{aligned}
I(2\omega)\propto P^{2}(2\omega)\propto|\chi_{a b c}^{(2)}|^{2}\propto|\chi_{a b c}^{(i)}\pm\chi_{a b c}^{(c)}|^{2},
\end{aligned}
\end{equation}  
where plus sign (+) corresponds to the positive magnetization direction whereas minus sign (-) represents the reversal of it. 

We investigate the polarization-resolved SHG intensity for all three directions (i.e. $x$, $y$ and $z$ direction) of incident light.
For normal incidence (i.e. along $z$ direction), the incident light polarization is in the $x$-$y$ plane. 
An azimuthal angle $\theta$ with respect to a reference direction, in this case the $x$ direction, 
can be used to describe the polarization direction of the incident light.
Following the {\it R}$3${\it c} symmetry of BiFeO$_3$, the response of in-plane ($P_{x}$ or $P_{y}$) 
and out-of-plane ($P_{z}$) SHG polarization for normal incidence can be written as,
\begin{equation}
\label{eq:12}
\begin{aligned}
& P_{x}=2\chi_{xxy}^{(2)}E_{x}E_{y} \\
& P_{y}=\chi_{yxx}^{(2)}E_{x}^{2}+\chi_{yyy}^{(2)}E_{y}^{2} \\
& P_{z}=\chi_{zxx}^{(2)}E_{x}^{2}+\chi_{zyy}^{(2)}E_{y}^{2},
\end{aligned}
\end{equation}
where $E_{x}$ and $E_{y}$ are the Cartesian components of the electric field of the incident light. 
The generated in-plane second harmonic light polarization direction can be either parallel or perpendicular 
to the polarization direction of the incident light.
Then, the parallel ($P_{\parallel}$) and perpendicular ($P_{\perp}$) components of the in-plane SHG polarization is defined as,
\begin{equation}
\label{eq:13}
\begin{aligned}
& P_{\parallel}=P_{x}cos\theta+P_{y}sin\theta \\
& P_{\perp}=-P_{x}sin\theta+P_{y}cos\theta,
\end{aligned}
\end{equation}
where $\theta$ is the azimuthal rotation angle.
By taking into account the symmetry-imposed shape of SHG susceptibility tensor (see Table I) 
and substituting them into Eqs. (\ref{eq:12})$\sim$(\ref{eq:13}), the SHG susceptibilities can be reduced to
\begin{equation}
\label{eq:14}
\begin{aligned}
& \chi_{\parallel}=\chi_{xxy}^{(2)}sin 3\theta \\
& \chi_{\perp}=\chi_{xxy}^{(2)}cos 3\theta \\
& \chi_{out}=\chi_{zxx}^{(2)}.
\end{aligned}
\end{equation}
Here $\chi_{\parallel}$ ($\chi_{\perp}$) are parallel (perpendicular) components of in-plane SHG susceptibilities 
whereas $\chi_{out}$ is the out-of-plane SHG susceptibility.
Then the SHG intensity would be proportional to $|\chi(\theta)|^{2}$.

Similarly, for light propagation along $x$ and $y$ direction, the polarization-resolved SHG susceptibilities can be written as,
\begin{equation}
\label{eq:15}
\begin{aligned}
& \chi_{\parallel}=-\chi_{xxy}^{(2)}cos^{3}\theta+(2\chi_{xxz}^{(2)}+\chi_{zxx}^{(2)})sin\theta cos^{2}\theta+\chi_{zzz}^{(2)}sin^{3}\theta \\
& \chi_{\perp}=\chi_{xxy}^{(2)}sin\theta cos^{2}\theta +(-2\chi_{xxz}^{(2)}+\chi_{zzz}^{(2)})sin^{2}\theta cos\theta+\chi_{zxx}^{(2)}cos^{3}\theta \\
& \chi_{out}=0,
\end{aligned}
\end{equation}
and
\begin{equation}
\label{eq:16}
\begin{aligned}
& \chi_{\parallel}=(2\chi_{xxz}^{(2)}+\chi_{zxx}^{(2)})sin^{2}\theta cos\theta+\chi_{zzz}^{(2)}cos^{3}\theta \\
& \chi_{\perp}=(2\chi_{xxz}^{(2)}-\chi_{zzz}^{(2)})sin\theta cos^{2}\theta-\chi_{zxx}^{(2)}sin^{3}\theta \\
& \chi_{out}=\chi_{xxy}^{(2)}sin^{2}\theta,
\end{aligned}
\end{equation} 
respectively.
$\theta$ is the azimuthal angle with respect to $y$($z$) direction for incident light along $x$($y$) direction, respectively. 

Figure \ref{fig:shg-z-in-pa} shows the parallel component of the calculated in-plane SHG intensity ($\propto|\chi_{\parallel}(\theta)|^2$) as a function of azimuthal polarization for normal incidence.
First, we notice that it exhibit a sixfold symmetry due to the sin3$\theta$ term. 
This is because of the threefold rotation symmetry in the $x$-$y$ plane.
Second, at SHG photon energy of 4.82 eV, we found a large change of 440\% in the SHG intensity with the reversal of the magnetization direction [see Fig. \ref{fig:shg-z-in-pa}(d)].
Also, from Figs. \ref{fig:shg-z-in-pa}(a)-\ref{fig:shg-z-in-pa}(c), at SHG photon energy of 2.9, 3.42 and 4.26 eV, the corresponding change in the SHG intensity is 196\%, 321\% and 273\%, respectively, when the magnetization direction is reversed.
We also notice that the relative change in the SHG intensity upon magnetization reversal is independent of azimuthal angle $\theta$.   
Furthermore, the perpendicular component of the in-plane SHG intensity ($\propto|\chi_{\perp}(\theta)|^2$) has a 30$^{\circ}$ rotation with respect to parallel component of the in-plane SHG intensity. 
Finally, for the out-of-plane SHG intensity ($\propto|\chi_{out}|^2$), both the SHG intensity as well as its relative change are independent of $\theta$. 

\begin{figure}[t] \centering
\includegraphics[width=8.0cm]{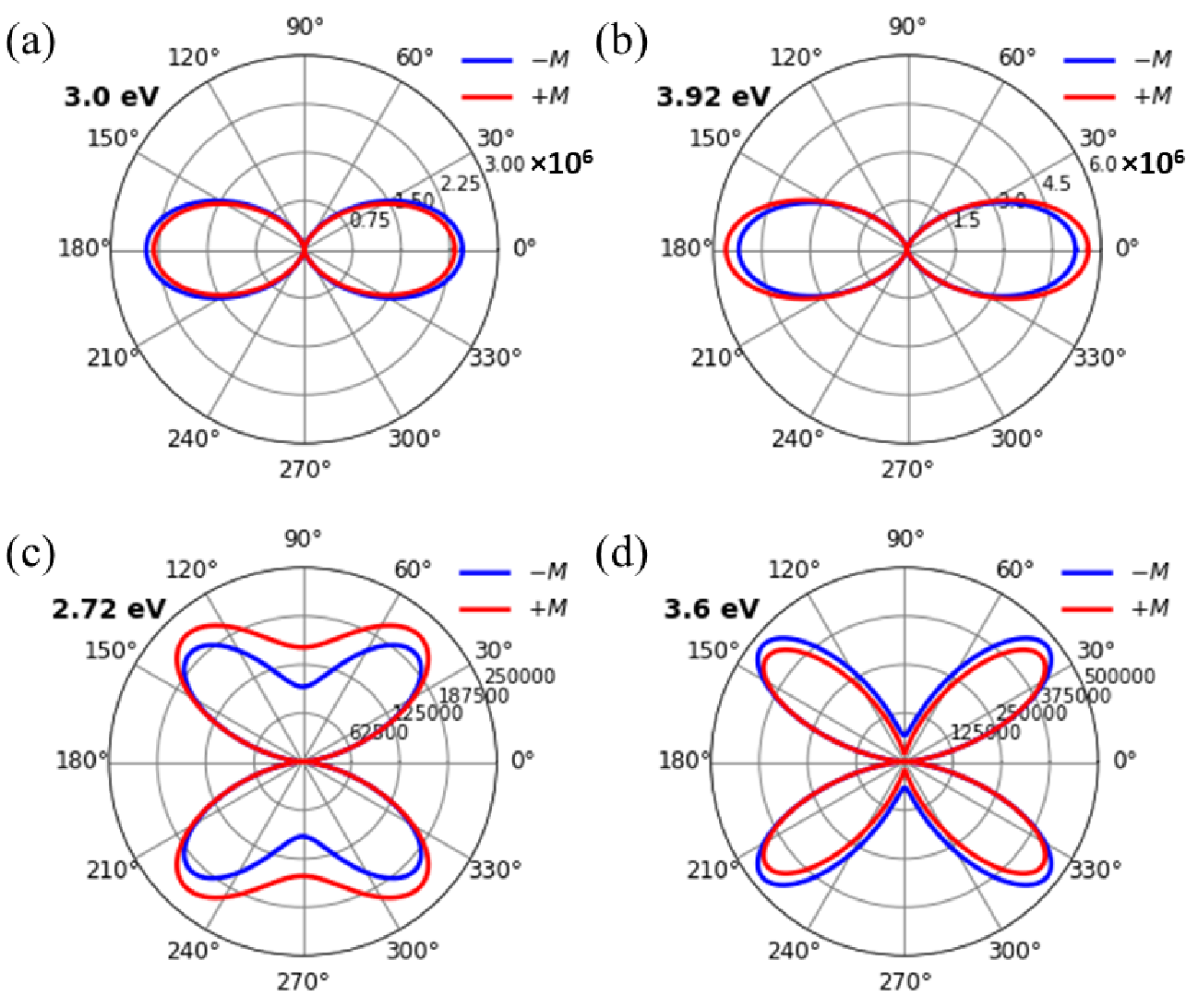}
\caption{Calculated in-plane polarization-resolved SHG pattern (a), (b) $|\chi_{\parallel}(\theta)|^2$, and (c), (d) $|\chi_{\perp}(\theta)|^2$ of BiFeO$_3$ as a function of azimuthal polarization for incident light along $y$-direction at SHG photon energy of (a) 3.0 eV, (b) 3.92 eV, (c) 2.72 eV, and (d) 3.6 eV.
Note that the red and blue spectra are for magnetization direction parallel and antiparallel to the $z$-axis, respectively.} 
\label{fig:shg-y-in-pa-pd}
\end{figure}

For the incident light along the $x$ direction, the out-of-plane SHG intensity vanishes.
The parallel [perpendicular] component of the in-plane SHG intensity for light propagation along the $x$ and $y$ direction 
are shown in Figs. \ref{fig:shg-x-in-pa-pd}(a)-\ref{fig:shg-x-in-pa-pd}(b) [Figs. \ref{fig:shg-x-in-pa-pd}(c)-\ref{fig:shg-x-in-pa-pd}(d)] and 
Figs. \ref{fig:shg-y-in-pa-pd}(a)-\ref{fig:shg-y-in-pa-pd}(b) [Figs. \ref{fig:shg-y-in-pa-pd}(c)-\ref{fig:shg-y-in-pa-pd}(d)], respectively.
Unlike the in-plane SHG intensity for normal incidence, here we find that both the in-plane SHG intensity 
as well as its relative change depends on $\theta$.
For instance, in the vicinity of azimuthal angle $\pi$ (or 2$\pi$), we find a maximum of 326\% of SHG intensity change 
upon magnetization reversal at SHG photon energy of 3.6 eV [see Fig. \ref{fig:shg-x-in-pa-pd}(d)].
For the light propagation along the $y$ direction, the maximum change in the in-plane SHG intensity goes up to 313\% 
(for $\theta$ = $\pi$/2 or 3$\pi$/2) when the magnetization is reversed [see Fig. \ref{fig:shg-y-in-pa-pd}(d)].
Upon the reversal of magnetization direction, the relative change in the out-of-plane SHG intensity for incident light 
along the $y$ direction follows the same relation as that of in-plane SHG intensity for normal incidence however 
it possess a two-lobed SHG pattern (see Figs. \ref{fig:shg-z-in-pa} and \ref{fig:shg-y-out}).
Interestingly, for incident light along $x$ (or $y$) direction, the parallel in-plane SHG intensity has two-lobed SHG pattern while the perpendicular component features a distorted butterfly-like pattern (see Figs.~\ref{fig:shg-x-in-pa-pd} and \ref{fig:shg-y-in-pa-pd}).

Thus, we find large magnetic contrast of the SHG signal of up to 440\% for SHG photon energy below 5 eV that results from the interference between $\chi_{a b c}^{(i)}$ and $\chi_{a b c}^{(c)}$. 
This suggests that the amplitude (or intensity) of SHG can be tuned via either electric or magnetic field.
Also, since the SHG signal responds sensitively to the reversal of the N{\'e}el vector, it may be used to detect the N{\'e}el vector reversal in antiferromagnetic materials, which is essential for antiferromagnetic spintronics.
Additionally, we show that the SHG susceptibilities are also strongly anisotropic in both shape and magnitude for different incident light directions.

\section{Magnetism-induced circular shift and linear injection current}

\begin{figure}[htbp] \centering
\includegraphics[width=8.0cm]{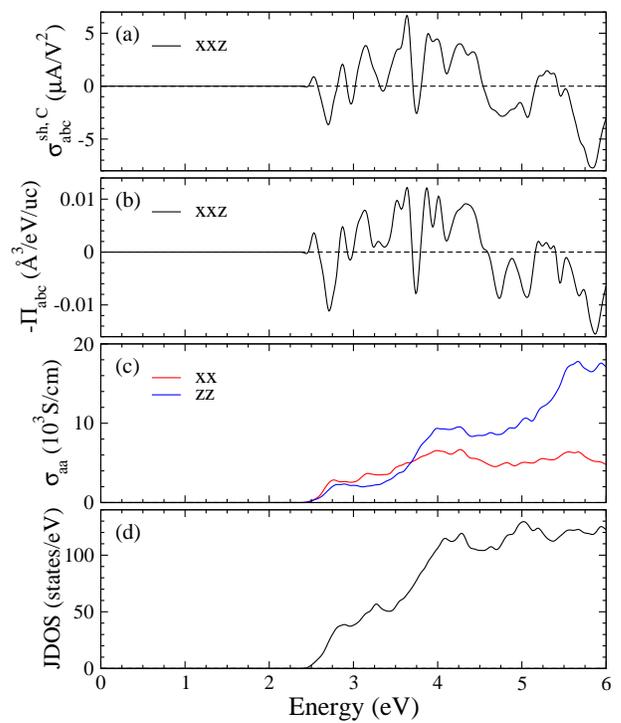}
\caption{(a) Circular shift current conductivity ($\sigma_{xxz}^{sh, C}$), 
(b) metric connection (-$\Pi_{xxz}$), 
(c) real part of optical conductivity ($\sigma_{xx}$ and $\sigma_{zz}$), 
and (d) joint density of states (JDOS) of BiFeO$_3$. 
In the metric connection unit, uc stands for unit cell.}
\label{fig:bpve-csc}
\end{figure}

As mentioned before, due to the breaking of both inversion and time-reversal symmetries,
antiferromagnetic BiFeO$_3$ possess all four types of BPVE, namely, linear and circular shift current conductivity
as well as linear and circular injection current susceptibility (Table I).
Like SHG susceptibility $\chi_{abc}^{(2)}$, both shift current conductivity ($\sigma_{abc}$) 
and injection current susceptibility ($\eta_{abc}$) are also a third-rank tensor, thus having 27 tensor elements.
However, because of the symmetry restrictions, only a few tensor elements are nonzero. 
In particular, Table I shows that circular shift current conductivity and injection current susceptibility
of BiFeO$_3$ have only one independent nonzero element (i.e., $\sigma_{xxz}^{sh, C}$ and $\eta_{xxz}^{inj, C}$),
while linear shift current conductivity and injection current susceptibility have four inequivalent nonzero elements.
Since the bulk photovoltaic effect of the crystallographic origin have been extensively studied 
before~\cite{Choi2009,Yang2009,Yang2010,Ji2010,Seidel2011,Ji2011,Alexe2011,Young2012b},
here we focus on the magnetism-induced circular shift and linear injection current.
Nevertheless, we present the results for the crystallographic linear shift and circular injection current
in the Supplementary Note 2 of the SM \cite{Prasad-SM}.

Magnetism-induced circular shift current conductivity ($\sigma_{xxz}^{sh, C}$) 
spectrum of BiFeO$_3$ is shown in Fig. \ref{fig:bpve-csc}(a).
Its visible spectral peak is at 2.70 eV for the negative maximum and 3.15 eV for the positive maximum, 
both of identical amplitude i.e. 4 ($\mu$A/V$^{2}$). 
The largest positive (negative) maximum, however, is 7 (-8) ($\mu$A/V$^{2}$) at 3.64 (5.84) eV.
Nevertheless, this largest negative maximum of $\sigma_{xxz}^{sh, C}$ is nearly $\sim$3 times 
smaller than the corresponding negative maximum of $\sigma_{xxz}^{sh, L}$ (see Fig. \ref{fig:bpve-csc}(a) 
and Fig. S5(a) in the SM \cite{Prasad-SM}). 
This shows that magnetism-induced $\sigma_{abc}^{sh, C}$ is smaller in magnitude compared with 
its nonmagnetic counterpart ($\sigma_{abc}^{sh, L}$) in BiFeO$_3$.

\begin{figure}[htbp] \centering
\includegraphics[width=8.0cm]{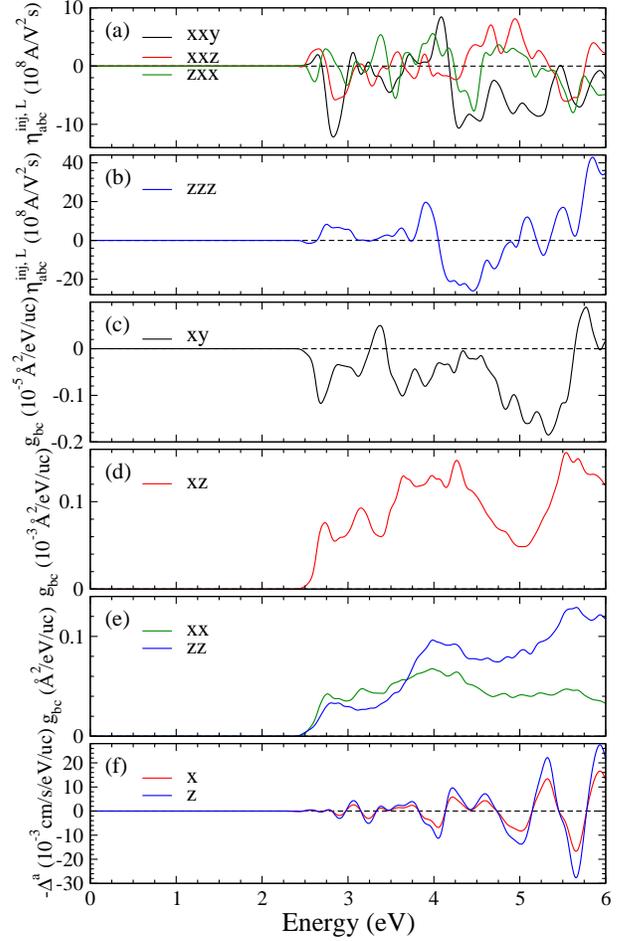}
\caption{(a-b) Linear injection current susceptibility ($\eta_{abc}^{inj, L}$), (c-e) quantum metric ($\textsl{g}_{bc}$), 
and (f) group velocity difference (-$\Delta^{a}$) of BiFeO$_3$. 
In quantum metric and group velocity difference unit, uc stands for unit cell.}
\label{fig:bpve-linj}
\end{figure}

The four spectra of linear injection current susceptibilities ($\eta_{abc}^{inj, L}$) are shown in Figs. \ref{fig:bpve-linj}(a)-\ref{fig:bpve-linj}(b). 
Figures \ref{fig:bpve-linj}(a) and \ref{fig:bpve-linj}(b) indicate that $\eta_{xxy}^{inj, L}$ 
and $\eta_{zzz}^{inj, L}$ have the largest negative and positive maximum of -12 $\times$ $10^{8}$ A/V$^{2}$s 
and $\sim$9 $\times$ $10^{8}$ A/V$^{2}$s at 2.83 and 2.75 eV, respectively, in the visible photon energy range.
$\eta_{xxy}^{inj, L}$ also has a positive maximum of $\sim$9 $\times$ $10^{8}$ A/V$^{2}$s at 4.09 eV [see Fig. \ref{fig:bpve-linj}(a)]. 
Moreover, $\eta_{zxx}^{inj, L}$ has a peak value of $\sim$6 $\times$ $10^{8}$ A/V$^{2}$s and -8 $\times$ $10^{8}$ A/V$^{2}$s 
at 3.38 and 4.47 eV, respectively. 
Figure \ref{fig:bpve-linj}(a) shows that $\eta_{xxz}^{inj, L}$ also has a positive maximum of 8 $\times$ $10^{8}$ A/V$^{2}$s at 4.94 eV. 
$\eta_{zzz}^{inj, L}$ has a broad positive peak 
of 20 $\times$ $10^{8}$ A/V$^{2}$s at 3.90 eV [Fig. \ref{fig:bpve-linj}(b)]. 
As $\hbar\omega$ increases, it decreases sharply, alters sign at 4.06 eV and then reaches to a largest negative maximum 
of -26 $\times$ $10^{8}$ A/V$^{2}$s at 4.45 eV.
In addition, $\eta_{zzz}^{inj, L}$ also has a large injection current susceptibility 
of 43 $\times$ $10^{8}$ A/V$^{2}$s at 5.85 eV [see Fig. \ref{fig:bpve-linj}(b)]. 

Furthermore, we notice that the positive maxima of $\eta_{zzz}^{inj, L}$ is two times larger
than $\eta_{xxz}^{inj, C}$ (see Fig. \ref{fig:bpve-linj}(b) and Fig. S6(a) in the SM \cite{Prasad-SM}).
Here it is important to note that, the positive maxima of $\eta_{zzz}^{inj, L}$ occur
at a higher photon energy (5.85 eV) compared to $\eta_{xxz}^{inj, C}$ ($\sim$4 eV).
In the vicinity of 4 eV, both $\eta_{zzz}^{inj, L}$ and $\eta_{xxz}^{inj, C}$ spectra have approximately the same magnitude.
Note that $\eta_{abc}^{inj, L}$ is purely due to magnetism.
Also, in comparison to $\eta_{xxz}^{inj, C}$ of CdS, the linear injection current
susceptibility $\eta_{zzz}^{inj, L}$ of BiFeO$_3$ is an order of magnitude larger.
Therefore, multiferroic BiFeO$_3$ could be a promising material for magnetic field-controllable photovoltaic
devices.

Interestingly, shift and injection current conductivities have recently been found to be related to 
the Hermitian connection and quantum geometric tensor of the electronic states, respectively \cite{Ahn2020,Hsu2023,Ahn2022}. 
Specifically, circular and linear shift current conductivities correspond to the real (metric connection) 
and imaginary (symplectic connection) parts of the Hermitian connection, respectively (see the SM~\cite{Prasad-SM} for details).
Linear and circular injection current conductivities are associated with the real (quantum metric) and imaginary (Berry curvature) 
parts of the quantum geometric tensor, respectively.
Therefore, to help understand the origins of the calculated photocurrent conductivities,
here we also display in Fig. \ref{fig:bpve-csc}(b) the photon energy-resolved metric connection (-$\Pi_{abc}$) 
[see Eq. (S11) in the SM~\cite{Prasad-SM}].
Indeed, Fig. \ref{fig:bpve-csc} shows that the -$\Pi_{xxz}$ spectrum looks very similar to the $\sigma_{xxz}^{sh, C}$ spectrum. 
In particular, the first small positive peak at 2.53 eV in $\sigma_{xxz}^{sh, C}$ corresponds to a positive peak in 
the -$\Pi_{xxz}$ spectrum.
Similarly, the negative peak of $\sigma_{xxz}^{sh, C}$ at 2.7 eV has a corresponding peak in the -$\Pi_{xxz}$ spectrum.
Thus, this demonstrates that all peaks (dips) in the $\sigma_{xxz}^{sh, C}$ spectrum
originate from the corresponding peaks in the metric connection spectra.
Furthermore, since the magnetism-induced linear injection current can be viewed as the product of group velocity difference 
and quantum metric (see Eq. 7), we present the photon energy-resolved spectra of quantum metric ($\textsl{g}_{bc}$) 
and group velocity difference (-$\Delta^{a}$) [see Eqs. (S13) and (S15) in the SM~\cite{Prasad-SM}] 
in Figs. \ref{fig:bpve-linj}(c)-\ref{fig:bpve-linj}(e) and \ref{fig:bpve-linj}(f), respectively.
We find that the two peaks of $\eta_{xxy}^{inj, L}$ at 3.06 and 5.69 eV have corresponding peaks 
in the -$\Delta^{x}$ spectrum [see Figs. \ref{fig:bpve-linj}(a) and \ref{fig:bpve-linj}(f)].
For $\eta_{xxz}^{inj, L}$, the first broad positive peak around 2.7 eV has a corresponding peak 
in $\textsl{g}_{xz}$ [see Figs. \ref{fig:bpve-linj}(a) and \ref{fig:bpve-linj}(d)], while the two negative peaks 
at approximately 2.9 and 3.28 eV can be attributed to the corresponding negative peaks in -$\Delta^{x}$ 
spectrum [see Figs. \ref{fig:bpve-linj}(a) and \ref{fig:bpve-linj}(f)].
Similarly, the three positive peaks of $\eta_{zxx}^{inj, L}$ at 2.74, 3.99, and 4.75 eV 
have corresponding peaks in the $\textsl{g}_{xx}$ spectrum [see Figs. \ref{fig:bpve-linj}(a) and \ref{fig:bpve-linj}(e)]. 
It also has a negative peak at 5.62 eV, corresponding to the negative peak in -$\Delta^{z}$ spectrum 
[see Figs. \ref{fig:bpve-linj}(a) and \ref{fig:bpve-linj}(f)].
Furthermore, the first positive peak of $\eta_{zzz}^{inj, L}$ has a corresponding peak in $\textsl{g}_{zz}$ 
near 2.75 eV [see Figs. \ref{fig:bpve-linj}(b) and \ref{fig:bpve-linj}(e)]. 
Also, at around 4 eV, both $\eta_{zzz}^{inj, L}$ and $\textsl{g}_{zz}$ spectra have a positive peak.
Thus, we can conclude that the peaks in the $\eta_{abc}^{inj, L}$ spectrum can be traced back to 
the prominent features in the quantum geometric quantities $\textsl{g}_{bc}$ and $\Delta^{a}$.

\section{CONCLUSIONS}
In summary, we have extended the Rashkeev {\it et al.}~\cite{Rashkeev1998}'s 
numerical calculation friendly formula for SHG to include magnetic systems, and then
systematically studied the 
magnetism-induced nonlinear optical responses in bismuth ferrite using first-principles DFT calculations.
Interestingly, we find that BiFeO$_3$ possess large nonlinear optical responses.
In particular, the calculated magnetism-induced SHG susceptibilities are large and 
the SHG intensity is tunable with magnetization (i.e., the N{\'e}el vector) reversal. 
At SHG photon energy close to 5 eV, we notice a significant magnetic contrast of the SHG signal of up to 440\% 
that results from the interference between crystallographic $i$-type SHG and magnetically induced $c$-type SHG.
This suggests that either electric or magnetic field can be used to tune the amplitude of SHG.
Because of their sensitivity to the N{\'e}el vector reversal, these SHG signals can be utilized to detect it 
in antiferromagnetic materials, which is necessary for antiferromagnetic spintronics.
Additionally, we show that the calculated SHG susceptibilities are also strongly anisotropic in both shape 
and magnitude for different incident light directions.
In terms of single-photon and double-photon resonances, the salient characteristics in the spectra 
of $\chi^{(2)}$ are also successfully associated with the features in the linear optical dielectric 
function $\varepsilon(\omega)$.
Furthermore, the calculated bulk photovoltaic responses of BiFeO$_3$ are also prominent and 
significantly larger than some of the popular NLO compounds (e.g. GaAs, CdS, CdSe and BaTiO$_3$).
Also, we have explained the origin of these pronounced peaks by comparing our calculated BPVE spectra 
with the corresponding quantum geometric quantities.
These interesting findings thus indicate that the magnetism-induced NLO responses of BiFeO$_3$ 
are pronounced in magnitude, anisotropic as well as tunable.
We believe that this work would stimulate further experimental and theoretical studies 
of magnetism-induced nonlinear optical responses in multiferroics.

\section*{ACKNOWLEDGMENTS}
The authors acknowledge the support from the National Science and Technology Council,
National Center for Theoretical Sciences, 
and the Academia Sinica in Taiwan. 
The authors also thank the National Center for High-performance Computing (NCHC) in Taiwan for the computing time.


\begin{thebibliography}{}
	
\bibitem{Boyd2003} R. W. Boyd, {\it Nonlinear Optics} (Elsevier, Amsterdam, 2003).

\bibitem{Shen2003} Y. R. Shen, {\it The Principle of Nonlinear Optics} (Wiley, Hoboken, NJ, 2003).

\bibitem{Franken1961} P. A. Franken, A. E. Hill, C. W. Peters, and G. Weinreich, 
Generation of Optical Harmonics, Phys. Rev. Lett. \textbf{7}, 118 (1961).

\bibitem{Kleinman1962} D. A. Kleinman, 
Theory of Second Harmonic Generation of Light, 
Phys. Rev. \textbf{128}, 1761 (1962).

\bibitem{Glass1974} A. M. Glass, D. von der Linde, and T. J. Negran, 
High-voltage bulk photovoltaic effect and the photorefractive process in LiNbO$_3$, 
Appl. Phys. Lett. \textbf{25}, 233 (1974).

\bibitem{Koch1975} W. T. H. Koch, R. Munser, W. Ruppel, and P. W{\"u}rfel, 
Bulk photovoltaic effect in BaTiO$_3$, 
Solid State Commun. \textbf{17}, 847 (1975).

\bibitem{Kraut1979} W. Kraut and R. von Baltz, 
Anomalous bulk photovoltaic effect in ferroelectrics: A quadratic response theory, 
Phys. Rev. B \textbf{19}, 1548 (1979).

\bibitem{Baltz1981} R. von Baltz and W. Kraut, 
Theory of the bulk photovoltaic effect in pure crystals, 
Phys. Rev. B \textbf{23}, 5590 (1981).

%
\bibitem{Fridkin2001} V. M. Fridkin, 
Bulk photovoltaic effect in noncentrosymmetric crystals, 
Crystallogr. Rep. \textbf{46}, 654 (2001).

\bibitem{Chang1965} R. K. Chang, J. Ducuing, and N. Bloembergen, 
Dispersion of the Optical Nonlinearity in Semiconductors, 
Phys. Rev. Lett. \textbf{15}, 415 (1965).

\bibitem{Levine1991a} Z. H. Levine and D. C. Allan, 
Calculation of the nonlinear susceptibility for optical second-harmonic generation in III-V semiconductors, 
Phys. Rev. Lett. \textbf{66}, 41 (1991).

%
\bibitem{Levine1991b} Z. H. Levine and D. C. Allan, 
Optical second-harmonic generation in III-V semiconductors: Detailed formulation and computational results, 
Phys. Rev. B \textbf{44}, 12781 (1991); 
Erratum: Optical second-harmonic generation in III-V semiconductors: Detailed formulation and computational results, \textbf{48}, 14768(E) (1993). 

%
\bibitem{Sipe1993} J. E. Sipe and Ed Ghahramani, 
Nonlinear optical response of semiconductors in the independent-particle approximation, 
Phys. Rev. B \textbf{48}, 11705 (1993).

\bibitem{Hughes1996} J. L. P. Hughes and J. E. Sipe, 
Calculation of second-order optical response in semiconductors, 
Phys. Rev. B \textbf{53}, 10751 (1996).

%
\bibitem{Cai2009} D. Cai and G.-Y. Guo, 
Tuning linear and nonlinear optical properties of wurtzite GaN by c-axial stress, 
J. Phys. D: Appl. Phys. \textbf{42}, 185107 (2009).

\bibitem{Cheng2019} M. Cheng, S. Wu, Z.-Z. Zhu, and G.-Y. Guo, 
Large second-harmonic generation and linear electro-optic effect in trigonal selenium and tellurium, 
Phys. Rev. B \textbf{100}, 035202 (2019).

\bibitem{Song2020a} W. Song, G.-Y. Guo, S. Huang, L. Yang, and L. Yang, 
First-principles Studies of Second-Order Nonlinear Optical Properties of Organic-Inorganic Hybrid Halide Perovskites, 
Phys. Rev. Appl. \textbf{13}, 014052 (2020). 
 
\bibitem{Guo2004} G. Y. Guo, K. C. Chu, D.-S. Wang, and C.-G. Duan, 
Linear and nonlinear optical properties of carbon nanotubes from first-principles calculations, 
Phys. Rev. B \textbf{69}, 205416 (2004).

\bibitem{Guo2005} G. Y. Guo and J. C. Lin, 
Second-harmonic generation and linear electro-optical coefficients of BN nanotubes, 
Phys. Rev. B \textbf{72}, 075416 (2005); 
Erratum: Second-harmonic generation and linear electro-optical coefficients of BN nanotubes, \textbf{77}, 049901(E) (2008).

\bibitem{Wu2008} I. J. Wu and G. Y. Guo, 
Second-harmonic generation and linear electro-optical coefficients of SiC polytypes and nanotubes, Phys. Rev. B \textbf{78}, 035447 (2008).

%
\bibitem{Wang2015} C.-Y. Wang and G.-Y. Guo, 
Nonlinear optical properties of transition-metal dichalcogenide MX$_2$ (M = Mo, W; X = S, Se) monolayers and trilayers from first-principles calculations, 
J. Phys. Chem. C \textbf{119}, 13268 (2015).

\bibitem{Hu2017} L. Hu, X. Huang, and D. Wei, 
Layer-independent and layer-dependent nonlinear optical properties of two-dimensional GaX (X = S, Se, Te) nanosheets, 
Phys. Chem. Chem. Phys. \textbf{19}, 11131 (2017).

\bibitem{Wang2017} H. Wang and X. Qian, 
Giant optical second harmonic generation in two-dimensional multiferroics, 
Nano Lett. \textbf{17}, 5027 (2017).

%
\bibitem{Attaccalite2019} C. Attaccalite, M. Palummo, E. Cannuccia, and M. Grüning, Second-harmonic generation in single-layer monochalcogenides: A response from first-principles real-time simulations, 
Phys. Rev. Mater. \textbf{3}, 074003 (2019).

\bibitem{Song2020b} W. Song, R. Fei, L. Zhu, and L. Yang, 
Nonreciprocal second-harmonic generation in few-layer chromium triiodide, 
Phys. Rev. B \textbf{102}, 045411 (2020).

\bibitem{Gudelli2020} V. K. Gudelli and G.-Y. Guo, 
Antiferromagnetism-induced second-order nonlinear optical responses of centrosymmetric bilayer CrI$_3$, 
Chin. J. Phys. \textbf{68}, 896 (2020).

\bibitem{Cheng2021} M. Cheng, Z.-Z. Zhu, and G.-Y. Guo, 
Strong bulk photovoltaic effect and second-harmonic generation in two-dimensional selenium and tellurium, 
Phys. Rev. B \textbf{103}, 245415 (2021). 

\bibitem{Gudelli2021} V. K. Gudelli and G.-Y. Guo, 
Large bulk photovoltaic effect and second-harmonic generation in few-layer pentagonal semiconductors PdS$_2$ and PdSe$_2$, 
New J. Phys. \textbf{23}, 093028 (2021).

\bibitem{Fiebig2005} M. Fiebig, V. V. Pavlov, and R. V. Pisarev, 
Second-harmonic generation as a tool for studying electronic and magnetic structures of crystals: review, 
J. Opt. Soc. Am. B \textbf{22}, 96 (2005).

\bibitem{Chen2022} H. Chen, M. Ye, N. Zou, B.-L. Gu, Y. Xu, and W. Duan, 
Basic formulation and first-principles implementation of nonlinear magneto-optical effects, 
Phys. Rev. B \textbf{105}, 075123 (2022). 

\bibitem{Xiao2022} R.-C. Xiao, D.-F. Shao, W. Gan, H.-W. Wang, H. Han, Z. G. Sheng, C. Zhang, H. Jiang, and H. Li, 
Classification of second harmonic generation effect in magnetically ordered materials, 
npj Quantum Mater. \textbf{8}, 62 (2023).

\bibitem{Toyoda2023} S. Toyoda, J.-C. Liao, G.-Y. Guo, Y. Tokunaga, T. Arima, Y. Tokura, and N. Ogawa, 
Magnetic-field switching of second-harmonic generation in noncentrosymmetric magnet  
Eu$_2$MnSi$_2$O$_7$, 
Phys. Rev. Mater. \textbf{7}, 024403 (2023).

\bibitem{Sipe2000} J. E. Sipe and A. I. Shkrebtii, 
Second-order optical response in semiconductors, 
Phys. Rev. B \textbf{61}, 5337 (2000).

\bibitem{Ahn2020} J. Ahn, G.-Y. Guo, and N. Nagaosa, 
Low-Frequency Divergence and Quantum Geometry of the Bulk Photovoltaic Effect in Topological Semimetals, 
Phys. Rev. X \textbf{10}, 041041 (2020).

\bibitem{Young2012a} S. M. Young and A. M. Rappe, 
First Principles Calculation of the Shift Current Photovoltaic Effect in Ferroelectrics, 
Phys. Rev. Lett. \textbf{109}, 116601 (2012).

\bibitem{Dai2022} Z. Dai and A. M. Rappe, 
Recent progress in the theory of bulk photovoltaic effect, 
Chem. Phys. Rev. \textbf{4}, 011303 (2023).

\bibitem{Zhang2019} Y. Zhang, T. Holder, H. Ishizuka, F. de Juan, N. Nagaosa, C. Felser, and B. Yan, 
Switchable magnetic bulk photovoltaic effect in the two-dimensional magnet CrI$_3$, 
Nat. Commun. \textbf{10}, 3783 (2019).

\bibitem{Fei2020} R. Fei, W. Song, and L. Yang, 
Giant photogalvanic effect and second-harmonic generation in magnetic axion insulators, 
Phys. Rev. B \textbf{102}, 035440 (2020).

\bibitem{Wang2020} H. Wang and X. Qian, 
Electrically and magnetically switchable nonlinear photocurrent in PT-symmetric magnetic topological quantum materials, 
npj Comput. Mater. \textbf{6}, 199 (2020).

\bibitem{Ma2017} Q. Ma, S.-Y. Xu, C.-K. Chan, C.-L. Zhang, G. Chang, Y. Lin, W. Xie, T. Palacios, H. Lin, S. Jia, P. A. Lee, P. Jarillo-Herrero, and N. Gedik, 
Direct optical detection of Weyl fermion chirality in a topological semimetal, 
Nat. Phys. \textbf{13}, 842 (2017).

\bibitem{Juan2017} F. de Juan, A. G. Grushin, T. Morimoto, and J. E. Moore, 
Quantized circular photogalvanic effect in Weyl semimetals, 
Nat. Commun. \textbf{8}, 15995 (2017).

\bibitem{Hsu2023} H.-C. Hsu, J.-S. You, J. Ahn, and G.-Y. Guo, 
Nonlinear photoconductivities and quantum geometry of chiral multifold fermions, 
Phys. Rev. B \textbf{107}, 155434 (2023).

\bibitem{Ahn2022} J. Ahn, G.-Y. Guo, N. Nagaosa, and A. Vishwanath, 
Riemannian geometry of resonant optical responses, 
Nat. Phys. \textbf{18}, 290 (2022).

\bibitem{Pi2023} H. Pi, S. Zhang, and H. Weng, 
Magnetic bulk photovoltaic effect as a probe of magnetic structures of EuSn$_2$As$_2$, 
Quantum Frontiers \textbf{2}, 6 (2023).

\bibitem{Butler2015}  K. T. Butler, J. M. Frost, and A. Walsh, 
Ferroelectric materials for solar energy conversion: Photoferroics revisited, 
Energy Environ. Sci. \textbf{8}, 838 (2015).

\bibitem{Tan2016}  L. Z. Tan, F. Zheng, S. M. Young, F. Wang, S. Liu, and A. M.
Rappe, 
Shift current bulk photovoltaic effect in polar materials—hybrid and oxide perovskites and beyond, npj Comput. Mater. \textbf{2}, 16026 (2016).

\bibitem{Cook2017} A. M. Cook, B. M. Fregoso, F. de Juan, S. Coh, and J. E. Moore, 
Design principles for shift current photovoltaics, 
Nat. Commun. \textbf{8}, 14176 (2017).

\bibitem{Sousa2016} R. de Sousa, The "Holy Grail" of Multiferroic Physics, 
Physics in Canada \textbf{72}, 57 (2016).

\bibitem{Neaton2005} J. B. Neaton, C. Ederer, U. V. Waghmare, N. A. Spaldin, and K. M. Rabe, First-principles study of spontaneous polarization in multiferroic BiFeO$_3$,
Phys. Rev. B \textbf{71}, 014113 (2005).

\bibitem{Ihlefeld2008} J. F. Ihlefeld, N. J. Podraza, Z. K. Liu, R. C. Rai, X. Xu, T. Heeg, Y. B. Chen, J. Li, R. W. Collins, J. L. Musfeldt, X. Q. Pan, J. Schubert, R. Ramesh, and D. G. Schlom, 
Optical band gap of BiFeO$_3$ grown by molecular-beam epitaxy, 
Appl. Phys. Lett. \textbf{92}, 142908 (2008). 

\bibitem{Yang2010} S. Y. Yang, J. Seidel, S. J. Byrnes, P. Shafer, C.-H. Yang, M. D. Rossell, P. Yu, Y.-H. Chu, J. F. Scott, J. W. Ager III, L. W. Martin, and R. Ramesh, 
Above-bandgap voltages from ferroelectric photovoltaic devices, 
Nat. Nanotechnol. \textbf{5}, 143 (2010). 

\bibitem{Grinberg2013} I. Grinberg, D. V. West, M. Torres, G. Gou, D. M. Stein, L. Wu, G. Chen, E. M. Gallo, A. R. Akbashev, P. K. Davies, J. E. Spanier, and A. M. Rappe, 
Perovskite oxides for visible-light-absorbing ferroelectric and photovoltaic materials, 
Nature \textbf{503}, 509 (2013). 

\bibitem{Kumar2008} A. Kumar, R. C. Rai, N. J. Podraza, S. Denev, M. Ramirez, Y.-H. Chu, L. W. Martin, J. Ihlefeld, T. Heeg, J. Schubert, D. G. Schlom, J. Orenstein, R. Ramesh, R. W. Collins, J. L. Musfeldt, and V. Gopalan, 
Linear and nonlinear optical properties of BiFeO$_3$, 
Appl. Phys. Lett. \textbf{92}, 121915 (2008). 

\bibitem{Ju2009} S. Ju, T.-Y. Cai, and G.-Y. Guo, 
Electronic structure, linear, and nonlinear optical responses in magnetoelectric multiferroic material BiFeO$_3$, 
J. Chem. Phys. \textbf{130}, 214708 (2009). 

%
\bibitem{Xu2023} S. Xu, J. Wang, P. Chen, K. Jin, C. Ma, S. Wu, E. Guo, C. Ge, C. Wang, X. Xu, H. Yao, J. Wang, D. Xie, X. Wang, K. Chang, X. Bai, and G. Yang, 
Magnetoelectric coupling in multiferroics probed by optical second harmonic generation, 
Nat. Commun. \textbf{14}, 2274 (2023). 

\bibitem{Choi2009} T. Choi, S. Lee, Y. J. Choi, V. Kiryukhin, and S.-W. Cheong, 
Switchable Ferroelectric Diode and Photovoltaic Effect in BiFeO$_3$, 
Science \textbf{324}, 63 (2009). 

\bibitem{Yang2009} S. Y. Yang, L. W. Martin, S. J. Byrnes, T. E. Conry, S. R. Basu, D. Paran, L. Reichertz, J. Ihlefeld, C. Adamo, A. Melville, Y.-H. Chu, C.-H. Yang, J. L. Musfeldt, D. G. Schlom, J. W. Ager III, and R. Ramesh, 
Photovoltaic effects in BiFeO$_3$, 
Appl. Phys. Lett. \textbf{95}, 062909 (2009). 

\bibitem{Ji2010} W. Ji, K. Yao, and Y. C. Liang, 
Bulk Photovoltaic Effect at Visible Wavelength in Epitaxial Ferroelectric BiFeO$_3$ Thin Films, 
Adv. Mater. \textbf{22}, 1763 (2010). 

\bibitem{Seidel2011} J. Seidel, D. Fu, S.-Y. Yang, E. A.-Lladó, J. Wu, R. Ramesh, and J. W. Ager III, 
Efficient Photovoltaic Current Generation at Ferroelectric Domain Walls, 
Phys. Rev. Lett. \textbf{107}, 126805 (2011). 
 
\bibitem{Ji2011} W. Ji, K. Yao, and Y. C. Liang, 
Evidence of bulk photovoltaic effect and large tensor coefficient in ferroelectric BiFeO$_3$ thin films, 
Phys. Rev. B \textbf{84}, 094115 (2011). 

\bibitem{Alexe2011} M. Alexe and D. Hesse, Tip-enhanced photovoltaic effects in bismuth ferrite, 
Nat. Commun. \textbf{2}, 256 (2011). 

\bibitem{Young2012b} S. M. Young, F. Zheng, and A. M. Rappe, 
First-Principles Calculation of the Bulk Photovoltaic Effect in Bismuth Ferrite, 
Phys. Rev. Lett. \textbf{109}, 236601 (2012).

\bibitem{Knoche2021} D. S. Knoche, M. Steimecke, Y. Yun, L. Mühlenbein, and A. Bhatnagar, 
Anomalous circular bulk photovoltaic effect in BiFeO$_3$ thin films with stripe-domain pattern, 
Nat. Commun. \textbf{12}, 282 (2021).

%
%
\bibitem{Kubel1990} F. Kubel and H. Schmid, 
Structure of a ferroelectric and ferroelastic monodomain crystal of the perovskite BiFeO$_3$,
Acta Cryst. B\textbf{46}, 698 (1990).

\bibitem{Perdew1996} J. P. Perdew, K. Burke, and M. Ernzerhof, Generalized Gradient Approximation Made Simple,
Phys. Rev. Lett. \textbf{77}, 3865 (1996); 
Erratum: Generalized Gradient Approximation Made Simple, \textbf{78}, 1396(E) (1997).

\bibitem{Blochl1994} P. E. Bl{\"o}chl, Projector augmented-wave method, 
Phys. Rev. B \textbf{50}, 17953 (1994).

\bibitem{Kresse1993} G. Kresse and J. Hafner, \textit{Ab initio} molecular dynamics for liquid metals, 
Phys. Rev. B \textbf{47}, 558 (1993).

\bibitem{Kresse1996} G. Kresse and J. Furthm{\"u}ller, Efficient iterative schemes for \textit{ab initio} 
total-energy calculations using a plane-wave basis set, 
Phys. Rev. B \textbf{54}, 11169 (1996).

\bibitem{Dudarev1998} S. L. Dudarev, G. A. Botton, S. Y. Savrasov, C. J. Humphreys, and A. P. Sutton, 
Electron-energy-loss spectra and the structural stability of nickel oxide: An LSDA+U study,
Phys. Rev. B \textbf{57}, 1505 (1998).

%
\bibitem{Jepson1971} O. Jepson and O. K. Anderson, The electronic structure of h.c.p. Ytterbium, Solid State Commun. \textbf{9}, 1763 (1971).

\bibitem{Temmerman1989} W. M. Temmerman, P. A. Sterne, G. Y. Guo, and Z. Szotek, Electronic Structure Calculations of High T$_{c}$ Materials,
Molecular Simulation \textbf{4}, 153 (1989).

%
\bibitem{Aversa1995} C. Aversa and J. E. Sipe, 
Nonlinear optical susceptibilities of semiconductors: Results with a length-gauge analysis, 
Phys. Rev. B \textbf{52}, 14636 (1995).

\bibitem{Rashkeev1998} S. N. Rashkeev, W. R. L. Lambrecht, and B. Segall, 
Efficient {\it ab initio} method for the calculation of frequency-dependent second-order optical response in semiconductors, 
Phys. Rev. B \textbf{57}, 3905 (1998).

\bibitem{Prasad-SM} See Supplemental Material at http://link.aps.org/supplemental/ for Figs. S1-S7, and supplementary Notes 1-3, 
which also include Refs. \cite{Ju2008,Nastos2010,Azpiroz2018}.

\bibitem{Ju2008} S. Ju and G.-Y. Guo, 
First-principles study of crystal structure, electronic structure, and second-harmonic generation in a polar double perovskite Bi$_2$ZnTiO$_6$, 
J. Chem. Phys. \textbf{129}, 194704 (2008).

%
%
%
\bibitem{Nastos2010} F. Nastos and J. E. Sipe, 
Optical rectification and current injection in unbiased semiconductors, 
Phys. Rev. B \textbf{82}, 235204 (2010).

%
\bibitem{Azpiroz2018} J. Ibañez-Azpiroz, S. S. Tsirkin, and I. Souza, 
{\it Ab initio} calculation of the shift photocurrent by Wannier interpolation, 
Phys. Rev. B \textbf{97}, 245143 (2018).

\bibitem{Nastos2006} F. Nastos and J. E. Sipe, 
Optical rectification and shift currents in GaAs and GaP response: Below and above the band gap, Phys. Rev. B \textbf{74}, 035201 (2006).

\bibitem{Yates2007} J. R. Yates, X. Wang, D. Vanderbilt, and I. Souza, 
Spectral and Fermi surface properties from Wannier interpolation, 
Phys. Rev. B \textbf{75}, 195121 (2007).
 
\bibitem{Marzari2012} N. Marzari, A. A. Mostofi, J. R. Yates, I. Souza, and D. Vanderbilt, 
Maximally localized Wannier functions: Theory and applications, 
Rev. Mod. Phys. \textbf{84}, 1419 (2012).

\bibitem{Pizzi2020} G. Pizzi \textit{et al.}, Wannier90 as a community code: new features
and applications, 
J. Phys.: Condens. Matter \textbf{32}, 165902 (2020).

%
%
%
%
\bibitem{Gallego2019} S. V. Gallego, J. Etxebarria, L. Elcoro, E. S. Tasci, and J. M. Perez-Mato, 
Automatic calculation of symmetry-adapted tensors in magnetic and non-magnetic materials: a new tool of the Bilbao Crystallographic Server, 
Acta Cryst. A\textbf{75}, 438 (2019).

%
%
\bibitem{Wadley2016} P. Wadley, B. Howells, J. \v{Z}elezn{\'y}, C. Andrews, V. Hills, R. P. Campion, V. Nov{\'a}k, K. Olejn{\'i}k, F. Maccherozzi, S. S. Dhesi, S. Y. Martin, T. Wagner, J. Wunderlich, F. Freimuth, Y. Mokrousov, J. Kune\v{s}, J. S. Chauhan, M. J. Grzybowski, A. W. Rushforth, K. W. Edmonds, B. L. Gallagher, and T. Jungwirth, 
Electrical switching of an antiferromagnet, 
Science \textbf{351}, 587 (2016).

\bibitem{Godinho2018} J. Godinho, H. Reichlov{\'a}, D. Kriegner, V. Nov{\'a}k, K. Olejn{\'i}k, Z. Ka\v{s}par, Z. \v{S}ob{\'a}\v{n}, P. Wadley, R. P. Campion, R. M. Otxoa, P. E. Roy, J. \v{Z}elezn{\'y}, T. Jungwirth, and J. Wunderlich, 
Electrically induced and detected Néel vector reversal in a collinear antiferromagnet, 
Nat. Commun. \textbf{9}, 4686 (2018).

\bibitem{Grigorev2021} V. Grigorev, M. Filianina, S. Yu. Bodnar, S. Sobolev, N. Bhattacharjee, S. Bommanaboyena, Y. Lytvynenko, Y. Skourski, D. Fuchs, M. Kläui, M. Jourdan, and J. Demsar, 
Optical Readout of the Néel Vector in the Metallic Antiferromagnet Mn$_2$Au, 
Phys. Rev. Appl. \textbf{16}, 014037 (2021).

\bibitem{Zhang2022} Y.-H. Zhang, T.-C. Chuang, D. Qu, and S.-Y. Huang, 
Detection and manipulation of the antiferromagnetic Néel vector in Cr$_2$O$_3$, 
Phys. Rev. B \textbf{105}, 094442 (2022).

%
%
%


\end{thebibliography}
\end{document}